\documentclass[twocolumn,prb,preprintnumbers,amsmath,amssymb]{aastex631} 
\usepackage{amsmath} 
\usepackage{amsfonts} 
\usepackage{comment}
\usepackage{graphicx}
\usepackage{dcolumn}
\usepackage{bm}
\usepackage{rotating}
\usepackage{color,soul}

\begin{document}

\title{Formation of Water-Rich Giant Planet Satellites at Decretion Disk Ice Lines}
\author{Teng Ee Yap \& David J. Stevenson}

\begin{abstract}
\indent  The volatile budgets of giant planet satellites are critical to unraveling the origin of their building blocks within the circumplanetary disks that hosted them. The Galilean moons Ganymede and Callisto, as well as the Saturnian moon Titan, are known to be anomalously water-rich on the basis of mean densities and interior models informed by gravity data from \textit{Galileo} and \textit{Cassini}, characterized by ice-to-rock ratios around unity. Here, we show that the water-ice sublimation line in a decreting circumplanetary disk lends itself to the formation of a water-rich solid reservoir, serving as a natural site for the birthplace of icy satellites. Fundamentally, this reflects how interior to the ice-line, water vapor is advected outwards, while beyond it, water-ice drifts inwards as pebbles. Using a semi-analytic model for dust and vapor evolution, we simulate vapor and ice accumulation at the ice-line, showing that solids just beyond the ice-line achieve steady-state ice-to-rock ratios a factor of a few higher than elsewhere in the disk. For typical disk parameters, this ice buildup occurs within a timescale of a few thousand years. We propose this as a first-order process that explains, at least to some extent, the compositions of three aforementioned satellites. We explore the impact of uncertain turbulence parameters on our results, namely the Schmidt number and Shakura-Sunyaev $\alpha$, before discussing them in the context of icy satellite D/H ratios. We conclude by evaluating alternative scenarios for explaining water-rich satellites, based on the conversion of CO to CH$_4$, with water as a byproduct. \\

\textit{Unified Astronomy Thesaurus concepts:} Galilean satellites (627), Saturnian satellites (1427), Titan (2186), Ganymede (2188), Callisto (2279)
\end{abstract}

\section{Introduction}
\indent Owing to their compositional and geological diversity, giant planet satellites have long been recognized as favorable sites for comparative planetary evolution, expanding the framework of possible planet formation outcomes with which we interpret the present plethora of discovered exoplanets. Most of the prominent, regular satellites (\textit{i.e.,} those on prograde, nearly circular and planar orbits) are understood to have formed in circumplanetary disks, themselves embedded within, and fed by, the circumsolar disk \textcolor{blue}{(Lubow et al., 1999; D'Angelo et al., 2002; Bate et al., 2003; Lunine \& Stevenson, 1982; Canup \& Ward, 2002; Mosqueira \& Estrada, 2003; Batygin \& Morbidelli, 2020; Krapp et al., 2024)}. Accordingly, their bulk compositions reflect the nature and provenance of their building blocks in such disks, shedding light on dust transport and processing therein. Their volatile inventories, in particular, constitute a first-order diagnostic of their origins with respect to sublimation fronts of principal ice species, most importantly water. Volatile abundances may, in addition, yield insights into the gas-phase chemistry that operated, and hence the thermodynamic conditions that prevailed, during their accretion. \\
\indent On the basis of mean densities, several giant planet satellites have been inferred to be anomalously water-rich, with ice-to-rock ratios around unity. These include the two outer Galilean moons, Ganymede and Callisto \textcolor{blue}{(Smith et al., 1979; Schubert et al., 1981)}, as well as the Saturnian moon Titan \textcolor{blue}{(Tyler et al., 1981)}. More precise determinations of the water-ice content in these moons have emerged from models of their interior structure, based largely upon gravity measurements from the Galileo and Cassini spacecrafts \textcolor{blue}{(\textit{e.g.,} Kuskov \& Kronrod, 2001; Sohl et al., 2002; Sohl et al., 2003)}, confirming their high water abundances. The observation that the three largest Solar System (SS) satellites happen to be the most water-rich, and roughly equally so, suggests a general process leading to water-ice enrichment at play in circumplanetary disk evolution. Indeed, the emergence of these water-rich compositions is not a trivial/obvious outcome. As we discuss below, considerations of circumsolar disk chemistry point to dust reservoirs with ice-to-rock ratios short of unity, and recent developments in our understanding of material infall onto circumplanetary disks indicate they, as a whole, are less water-rich than the circumstellar disks they inhabit. 

\subsection{Expected Solar Ice-to-Rock Ratio}
\indent An ice-to-rock ratio around unity is peculiar given our understanding of the ``available" oxygen present for water-ice production in the circumsolar disk. On the basis of photospheric spectra and analysis of CI chondrites, the solar C/O ratio is estimated to be $\sim 0.5$, and $\sim 23\%$ of all oxygen atoms are estimated to be dedicated to rock condensation \textcolor{blue}{(\textit{e.g.,} Lodders, 2003; Asplund et al., 2009; Caffau et al., 2011)}. Given that most of the carbon in molecular clouds and the interstellar medium (ISM) take the form of CO \textcolor{blue}{(\emph{e.g.,} Lacy et al., 1991; Herbst, 1995)}, say with CO/CH$_{4}$ $\sim1\%$, this amounts to another (absolute) reduction of $\sim 50\%$ in the total available oxygen. This leaves $\sim 27\%$ of oxygen atoms for H$_{2}$O, assuming no CO is converted to CO$_2$. The validity of this assumption depends on the dominant mode of carbon gas-phase chemistry. \\
\indent In a H$_2$-dominated medium, this chemistry is governed by the reaction  \textcolor{blue}{(\textit{e.g.,} Prinn \& Barshay, 1977; Prinn \& Fegley, 1981; Lodders, 2003)}:
\begin{equation}
CO + 3H_{2} \rightleftharpoons CH_{4} + H_{2}O.
\end{equation}
\indent The disposition of carbon in this reaction is set by the chemical potentials of the species (which depend on their partial pressures, among other factors), whence a prediction can be made for their relative abundances. With four molecules on the left of the reaction, and only two on the right, high pressure favors the production of CH$_4$ and H$_2$O by the law of mass action. Accordingly, the total pressure is important in deciding the reaction outcome. The result is that the predicted fraction $f_O$ of oxygen in \textbf{Eq. 1} in the form of H$_2$O is given by
\begin{equation}
    f_O \approx 1/(1+P_c^2/P^2),
\end{equation} 
where $P_c$ is dependent on the quench temperature ($T_Q \sim 1000K$; \textcolor{blue}{Cavalie et al., 2023}) and on the order of $\sim 0.1$ bar. Gas pressures in both circumstellar and circumplanetary disks typically range between $10^{-4}$ and $10^{-3}$ bar, such that the CO-CH$_4$ conversion (and thus H$_2$O production) is kinetically inhibited. In this case, carbon chemistry is dominated by either graphitization, given by
\begin{equation}
2CO \rightleftharpoons C_{Graphite} + CO_2,
\end{equation}
or oxidation with available H$_2$O \textcolor{blue}{(Lodders, 2003)}, given by
\begin{equation}
CO + H_{2}O \rightleftharpoons CO_2 + H_{2}.
\end{equation}
\indent Assuming $\sim20\%$ of available CO is converted to CO$_2$ by the above reaction leads to an additional (absolute) reduction of $\sim10\%$ in the total available oxygen, leaving only $\sim17\%$ for H$_2$O. \\
\indent With (i) an ice-to-rock ratio (by mass) of unity (corresponding to a ice volume fraction of $\rho_{rock}/(\rho_{rock}+\rho_{ice}) \sim 70\%$; $\rho_{ice}\simeq 1.2$ g/cc and $\rho_{rock} \simeq 3$ g/cc), and (ii) assuming forsterite (Mg$_2$SiO$_{4}$) as the dominant silicate phase, the ratio of oxygen held in ice to that in rock within Ganymede, Callisto, and Titan is $\sim1.9$. For comparison, the expectation from partitioning of oxygen atoms is $\sim 17/23\sim 0.74$ (\textit{i.e.,} ice-to-rock ratio $\sim 0.4$). Thus, water-ice abundances in the three satellites are super-solar by a factor a few. Note that if no CO-CO$_2$ conversion is assumed, the said ratio expected would be $\sim 27/23\sim 1.2$ (\textit{i.e.,} ice-to-rock ratio $\sim 0.6$), still less than that observed. Moreover, we have neglected the potential for post-accretional volatile loss induced by impacts and/or tidal heating. That is, the feedstock for satellite accretion may have been even more water-rich than the satellites currently are. 

\subsection{Dry Dust Delivery into Circumplanetary Disks}
\indent In addition to disk chemistry, our current perception of dust coagulation and material delivery into circumplanetary disks (from their parent circumstellar disks) suggests the three said satellites ought to be more rocky than they are. Hydrodynamic simulations and astronomical probes (\textit{i.e.,} via $^{12}$CO emission) of gas flow around giant planets undergoing runaway gas accretion indicate gas and dust delivery into circumplanetary disks are facilitated through meridional flows originating from \textit{approximately one hydrostatic scale height $H$ above the circumstellar disk midplane} \textcolor{blue}{(Tanigawa et al., 2012; Morbidelli et al., 2014; Teague et al., 2019; Szulágyi et al., 2022)}. By virtue of a balance between gravitational settling and turbulent diffusion, dust particles of a given size are known to form a sub-disk, with scale height $H_d$ given by \textcolor{blue}{(Dubrulle et al., 1995)}
\begin{equation}
H_d = H\left(1+\frac{St}{\alpha}\right)^{-1/2},
\end{equation}
where $St$ represents the dimensionless particle Stokes number, directly related to particle size, and $\alpha$ the well-known Shakura-Sunyaev turbulence parameter \textcolor{blue}{(Shakura \& Sunyaev, 1973)}. For a typical $\alpha$ value of $\sim 10^{-3}$ for the circumsolar disk \textcolor{blue}{(\textit{e.g.,} Bitsch et al., 2015; Dullemond et al., 2018; Rosotti et al., 2023; Yap et al., 2025)}, it is clear that only particles with $St\lesssim $ few times $10^{-4}$ (\textit{i.e.,} mm-sized or less for typical disk parameters) will have $H_d\simeq H$ and thus be subsumed into circumplanetary disks. \\
\indent In dust coagulation models, collisional fragmentation due to relative turbulent velocities is typically identified as the barrier to hit-and-stick particle growth \textcolor{blue}{(\textit{e.g.,} Birnstiel et al., 2011, 2012; Batygin \& Morbidelli, 2022; Yap \& Batygin, 2024)}. These models are informed by laboratory experiments on dust collisions in microgravity, which indicate that icy particles are characterized by higher surface energies, and thus fragmentation threshold velocities, than purely rocky particles (\textit{i.e.,} those within the water-ice sublimation line; henceforth denoted the ``ice-line"; \textcolor{blue}{Blum \& Münch, 1993; Güttler et al., 2010; Gundlach et al., 2011; Gundlach \& Blum, 2014}). As such, icy particles can grow more efficiently and larger (as turbulent relative velocities $\Delta v_t \sim \sqrt{St}$; \textcolor{blue}{Ormel \& Cuzzi, 2007}), and thus settle more readily, than rocky particles. Assuming the water-ice fraction of particles lofted to height $\sim H$ is smaller than that at the midplane (more ice is held in larger, more settled particles), then, leads to the expectation that dust delivered into circumplanetary disk and hence satellite formation regions therein ought to be sub-solar. Regarding the Jovian disk in particular, Jupiter is widely thought to have accreted just beyond the circumsolar disk ice-line \textcolor{blue}{(\textit{e.g.,} Morbidelli et al., 2022)}. As meridional flows deliver gas and dust onto circumplanetary disks from both sides of the giant planet orbit, this suggests a substantial portion of dust subsumed into the Jovian disk originated from within the ice-line, and was thus volatile-depleted.

\begin{figure} 
\scalebox{1.5}{\includegraphics{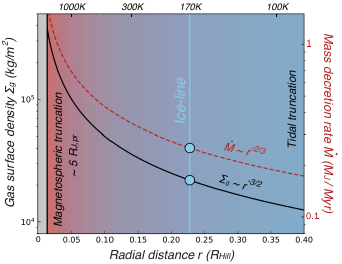}}
\caption{\textbf{Gas surface density $\Sigma_g(r)$ (black) and mass decretion rate $\dot{M}(r)$ (red) profiles in our circumplanetary (Jovian) disk model ($\alpha = 10^{-4}$).} The disk inner edge at $\sim 5$ times Jupiter's (primordial) radius $R_{J,pr} \simeq 2 R_J \simeq 1.4\times 10^8$m \textcolor{blue}{(Batygin \& Adams, 2025)} is set by magnetospheric truncation. The disk outer edge at $\sim0.4 R_{Hill}$ is set by tidal truncation \textcolor{blue}{(Martin \& Lubow, 2011)} from perturbations by the (proto-)Sun. The ice-line at $T(r_i)=170K$ is situated at $r_i\simeq 0.23 R_{Hill}$, where $\Sigma_g(r_i)\simeq 2.2\times 10^4$ kg/m$^2$ and $\dot{M}\simeq 0.25M_J/Myr$. See text in \textbf{Section 2.1} for details.}
\label{fig:Figure 1}
\end{figure}

\subsection{Proposed Scenario for Water-Ice Enrichment}
\indent In this work, we propose a ``cold finger" mechanism for explaining the high water-ice abundance in giant planet satellites. Relying on a semi-analytic vapor and dust evolution model, we show that the currently favored paradigm of decreting circumplanetary disks naturally lends itself to rapid (\textit{i.e.,} sub-10-kyr) ice accumulation at the ice-line, and thus the establishment of a water-rich solid reservoir for satellite accretion. In our model, the final ice-to-rock ratio of the solid reservoir reflects the competition between the outward advection of dust and water vapor with disk gas, vapor diffusion, and inward pebble drift, and indicates the formation of water-rich satellitesimals is possible if not inevitable. We make no attempt to identify exact conditions leading to the formation of Ganymede, Callisto, and Titan, aiming largely at bringing attention to what we believe is a universal physical process. That is, we pursue this work in the spirit of the maxim: "focus on \textit{process}, not \textit{scenario}."\\
\indent The paper is structured as follows: in \textbf{Section 2}, we describe our circumplanetary disk model, as well as our treatment of dust and vapor evolution at the ice-line. In \textbf{Section 3}, we show that for a reasonable (or conservative) set of model parameters (\textit{e.g.,} disk metallicity, initial ice content), accumulation of solids with an ice-to-rock ratio close to unity can be achieved in a timescale on the order of a few thousand years. Here, we also explore the impact of highly uncertain turbulence parameters, namely the Schmidt number and $\alpha$, on our results. In \textbf{Section 4}, we discuss our results in the context of icy staellite D/H ratios before evaluating the potential for CO-CH$_4$ conversion (\textbf{Eq. 1}) to explain water-rich satellites. Our concluding remarks are provided in \textbf{Section 5}. 

\section{Model Description}

\subsection{The Circumplanetary Decretion Disk}
\indent Meridional flows delivering gas and dust into circumplanetary disks are thought to do so in the vicinity of their inner edges, carved by magnetospheric truncation \textcolor{blue}{(\textit{e.g.,} Ghosh \& Lamb, 1979; Ostriker \& Shu, 1995; Mohanty \& Shu, 2008)}, and result in gas \textit{decretion} across most of their extents. This includes our region of interest (\textit{i.e.,} the ice-line). Here, we adopt a model for a viscous decretion disk, using the $\alpha$-prescription for turbulent viscosity \textcolor{blue}{(Shakura \& Sunyaev, 1973)} and assuming Jupiter as the giant planet. Our model resembles the ``gas-starved" disk of \textcolor{blue}{Canup \& Ward (2002)}, in that gas and dust is envisioned to be supplied and cycled through the disk across its lifetime.\\
\indent We assume a power surface density profile $\Sigma_g(r)$ of index $\gamma$, taking the simple form 
\begin{equation}
\Sigma_g(r) = \Sigma_0\left(\frac{r}{r_0}\right)^{-\gamma},
\end{equation}
and anchored to $\Sigma_0 =  5\times 10^4$ kg/m$^2$ at radial distance $r_0= 0.1 R_{Hill}$ \textcolor{blue}{(Batygin \& Morbidelli, 2020)}, where $R_{Hill}$ is the giant planet (Jovian) Hill radius, given by 
\begin{equation}
R_{Hill} = a_J\left(\frac{M_J}{3M_{\odot}}\right)^{1/3}.
\end{equation}
Here, $a_J \simeq 7.8\times 10^{11}$m is Jupiter's semi-major axis, $M_J \simeq 1.9\times 10^{27}$ kg Jupiter's mass, and ${M}_\odot \simeq 10^3 M_J$ the solar mass. We assume $\gamma=1$ \textcolor{blue}{(Adams \& Batygin, 2025)}, a value which, as shown below, renders our model consistent with net decretionary gas flow. The turbulent viscosity facilitating disk evolution takes the form
\begin{equation}
\nu = \alpha c_s h = \alpha \frac{c_s^2}{\Omega_k},
\end{equation}
where $c_s = \sqrt{k_b T(r)/\mu}$ represents the isothermal sound speed, with $T(r)$ being the disk midplane temperature, $k_b\simeq 1.38\times 10^{-23}$ m$^2\cdot$ kg$\cdot$ s$^{-2}\cdot$ K$^{-1}$ the Boltzmann constant, and $\mu\simeq 2.4$ proton masses $\simeq 4\times 10^{-27}$ kg the mean molecular mass for the H$_2$/He disk gas. The hydrostatic scale height $h \simeq c_s/\Omega_k$, where the Keplerian angular velocity $\Omega_k = \sqrt{G M_J/r^3}$, with the gravitational constant $G=6.67\times10^{-11}$ m$^3\cdot$ kg$^{-1}\cdot$ s$^{-2}$. Though highly uncertain, we assume a fiducial value of $\simeq 10^{-4}$ for the $\alpha$ turbulence parameter.\\
\indent For simplicity, we assume a disk heated solely by viscous dissipation for which the temperature profile $T(r)$ takes the form \textcolor{blue}{(\textit{e.g.,} Armitage, 2020; Yap \& Batygin, 2024)} 
\begin{equation}
T(r) = \left(\frac{27\mathbb{Z_\mu}\kappa_d\alpha k_b\Omega_k\Sigma_g(r)^2}{64\sigma\mu}\right)^{1/3},
\end{equation}
where $\mathbb{Z}_\mu$ is the ``micron-scale" metallicity accounting for opacity, $\kappa_d \simeq 30$ kg/m$^2$ the dust opacity \textcolor{blue}{(Bitsch et al., 2014)}, and $\sigma\simeq 5.67\times 10^{-8}$ W$\cdot$ m$^{-2}\cdot$ K$^{-4}$ the Stefan-Boltzmann constant. With $\mathbb{Z}_{\mu} = 5\times10^{-4}$ (\textit{i.e.,} $\sim 5\%$ of the typical total ISM/solar metallicity $\mathbb{Z}_{ISM} = 1\%$; \textcolor{blue}{\textit{e.g.,} Bohlin et al., 1978}), the ice-line ($T\simeq 170K$) is positioned at $r_i \simeq 0.23 R_{Hill}$. \\
\indent For a geometrically thin disk, continuity of mass and angular momentum yields a gas radial velocity given by \textcolor{blue}{(Armitage, 2020)} 
\begin{equation}
v_{g,r} = -\frac{3}{\Sigma_g\sqrt{r}}\frac{d}{dr}\left(\nu\Sigma_{g}\sqrt{r}\right), 
\end{equation}
Clearly, for viscous exchange between disk annuli to result in decretionary/outward net flow (\textit{i.e.,} for $v_{g,r}$ to be positive), $\nu\Sigma_{g,r}$ must be a function that falls more steeply than $\sqrt{r}$. \\
\indent Whether or not $v_{g,r}>0$ is satisfied depends on the value of $\gamma$ from \textbf{Eq. 6}. Substituting \textbf{Eqs. 6, 8, \& 9} into \textbf{Eq. 10}, we find
\begin{equation}
v_{g,r} = \left(5\gamma - \frac{9}{2}\right)\frac{\alpha c_s^2}{v_k},
\end{equation}
where $v_k = r\Omega_k$ is the Keplerian orbital velocity. The mass decretion rate $\dot{M} = 2\pi r\Sigma_g v_{g,r}$ then takes the form
\begin{equation}
\dot{M} = \pi(10\gamma-9)\frac{\alpha c_s^2 \Sigma_0r_0^\gamma}{v_k r^{\gamma-1}}.
\end{equation}
Thus, only for $\gamma>0.9$ is gas flow directed radially outwards, and our adopted value of $\gamma = 1$ is in accordance with this result. For our aforementioned choice of fiducial parameters, $\dot{M}(r)\sim r^{1-5\gamma/3}\sim r^{-2/3}$, and evaluates to $\sim0.25 M_J/Myr$ at the ice-line.\\
\indent While a decretion disk is not, technically speaking, in steady-state, it behooves us to consider if it can be assumed to be\textemdash that is, if the timescale for the evolution of $\Sigma_g(r)$ is comparable or greater than that of ice accumulation. In that case, $\Sigma_g(r)$ and $T(r)$ can be safely assumed to remain fixed, greatly simplifying our treatment of dust evolution to come. The evolution of $\Sigma_g(r)$ can be expressed as
\begin{equation}
\dot{\Sigma_g} \sim -\frac{1}{r}\frac{d\dot{M}}{dr} = \left(\frac{5\gamma}{3}-1\right)\frac{\dot{M}}{r}.
\end{equation}
From this, we can derive a timescale $\tau_{\Sigma} \sim \Sigma_g/\dot{\Sigma_g}$ in which $\Sigma_g$ changes appreciably, yielding
\begin{equation}
\tau_{\Sigma} \sim \frac{(\Sigma_0r_0^\gamma)r^{2-\gamma}}{(1-5\gamma/3)\dot{M}}.
\end{equation}
For $\gamma= 1$, this evaluates to $\simeq 10^4$ yrs at $r_i$.  We confirm \textit{a posteriori} that the timescales of interest to our work, namely the time it takes for the ice-to-rock ratio and dust metallicities in our model to stabilize, falls short of $\tau_\Sigma$ by at least a factor of a few. Hence, steady-state can be safely assumed.

\subsection{Qualitative Picture for Ice Accumulation}
\indent Before developing a quantitative treatment for dust evolution at the ice-line, we ought to consider qualitatively the reason we expect it to facilitate water-ice accumulation. The ice-line in accretion disks has long been recognized as a region promoting dust/ice buildup \textcolor{blue}{(\textit{e.g.,} Stevenson \& Lunine, 1988; Cuzzi \& Zahnle, 2004; Ciesla \& Cuzzi, 2006; Kretke \& Lin, 2007; Brauer et al., 2008a; Ros \& Johansen, 2013; Drażkowska \& Alibert, 2017; Schoonenberg \& Ormel, 2017; Morbidelli et al., 2022)}. This is thought to result from a ``cold finger" effect, whereby ice sublimes from the surfaces of inward-drifting pebbles, and diffuses back across the ice-line, re-condensing into and onto solid pebbles (note, however, that the direction of vapor advection is inward, away from the ice-line). Moreover, the drift of icy pebbles past the ice-line has been shown to result in a  ``traffic jam" effect, as rocky pebbles freed of their ice once they past the ice-line experience slower drift owing to their smaller size  \textcolor{blue}{(Drażkowska \& Alibert, 2017)}. This is amplified by the expected lower fragmentation threshold of rocky pebbles relative to their icy counterparts (see \textbf{Section 1.3}), limiting growth once within the ice-line.\\
\indent Ice accumulation via vapor diffusion across the ice-line in a Jovian accretion disk has recently been proposed to explain high water abundances in the formation region of Ganymede and Callisto \textcolor{blue}{(Mousis et al., 2023)}. In this scenario, water vapor released from dehydrated phyllosilicates (\textit{i.e.,} hydrous minerals) within the ice-line (at $T\gtrsim 400K$) is diffused beyond it. Notably, the authors show that an ice-to-rock ratio close to unity can be reproduced within $\sim 0.1$ Myr of disk evolution (this, as shown below, is approximately an order of magnitude longer than it takes for such a ratio to be achieved in our model; see \textbf{Section 3} below). \\
\indent To the best of our knowledge, ice accumulations at ice-lines in \textit{decretion disk}s remains absent in the literature. Nonetheless, dust accumulation at the silicate sublimation line during early decretionary phases of circumstellar disks has been proposed to explain the formation of planetesimal rings, and ultimately the ubiquitous ``peas-in-a-pod" architecture of extrasolar systems within the \textit{Kepler} survey \textcolor{blue}{(Weiss et al., 2018; Batygin \& Morbidelli, 2023)}. In these works, dust accumulation occurs as silicate vapor advected past the sublimation line condenses into pebbles that then drift back towards it. \\
\indent In decretion disks, an increase in the ice-to-rock ratio at the ice-line is driven by \textit{both advection and diffusion} of water vapor interior to it. This vapor is sourced from both the inner disk (\textit{i.e.,} ``fresh" infall) as well as the sublimation of inward drifting icy pebbles from the outer disk. Consider the following sequence of events, depicted in \textbf{Fig. 2}: (i) Small rocky particles (tightly coupled to the decreting gas) along with water vapor are advected across the ice-line, resulting in the condensation of water-ice onto the said pebbles; (ii) the icy pebbles, now larger (they also undergo collisional growth), drift back across the ice-line, resulting in sublimation and the release of rocky particles; while within the ice-line, these particles can grow, such that their outward advection is slowed; (iii) over several iterations of (i) and (ii), the rocky particles grow large enough to drift \textit{away} from the ice-line. The water, however, is largely preserved in going back and forth across the ice-line, thereby building up over time. Note that (i) and (ii) can be interchanged, depending on whether the vapor and rocky particles are initially sourced from the inner or outer disk. In the latter case, it is conceivable that a fraction of rocky particles released by icy sublimation are already sufficiently large to proceed in drifting towards the giant planet. \\
\indent All in all, as water (vapor) within the ice-line is advected outwards towards it, and water (ice) beyond the ice-line largely drifts inwards towards it, ice inevitably accumulates thereat. There are only two (local) sinks: the advection and diffusion of small icy particles farther out in the disk and the diffusion of water vapor back towards the inner disk. The said particles, if not lost to the circumsolar disk, will eventually grow and drift back towards the ice-line. Likewise, absent of a strong gradient in vapor content to drive diffusion deeper into the inner disk, the said vapor will eventually be advected back towards the ice-line. 

\begin{figure} 
\scalebox{1.7}{\includegraphics{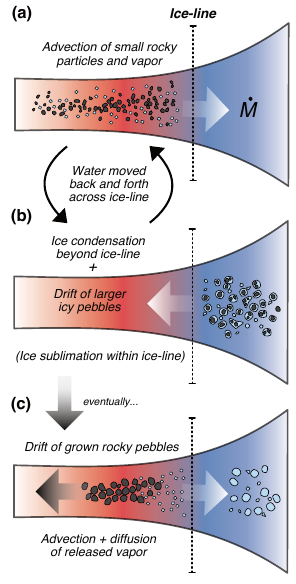}}
\caption{\textbf{Qualitative sketch of water-ice accumulation at the decretion disk ice-line.} \textbf{(a)} Small and tightly coupled rocky particles, along with water vapor, are advected across the ice-line with disk gas. \textbf{(b)} Beyond the ice-line, water-ice condenses onto the rocky particles, forming larger icy particles that drift back across the ice-line. \textbf{(c)} Over many iterations of (interchangeable) steps \textbf{(a)} and \textbf{(b)}, the rocky particles grow among themselves, such that they eventually drift back to the giant planet. The water that originally accompanied these particles remains at the ice-line, as released vapor is continually advected (and diffused) outwards. See text in \textbf{Section 2.2}.}
\label{fig:Figure 2}
\end{figure}

\subsection{Semi-Analytic Dust Evolution Model at Ice Line}
\indent To quantitatively illustrate the envisioned scenario for ice buildup, we develop a semi-analytic, ``two-bin" model for dust evolution at the ice-line. The two bins represent two disk annuli of radial width $\Delta r$, separated by the ice-line at a distance $r_i$ from the giant planet. We set $\Delta r = 0.01 R_{Hill} \ll r_i = 0.23R_{Hill}$, such that the area of each bin is $A\simeq 2\pi r_i\Delta r$. The first bin ($r_i-\Delta r$ to $r_i$) hosts gas, silicate dust, and water vapor, while the second ($r_i$ to $r_i + \Delta r$) hosts gas and icy dust. We adopt a two-population model for dust particles \textcolor{blue}{(Birnstiel et al., 2012)}, wherein the budget of solids across the disk is treated as a bimodal mixture of (i) micron-sized particles (contributing to the disk opacity) and (ii) larger particles (henceforth referred to as ``pebbles") with a characteristic size in the mm- to cm-scale. \\
\indent Gas and dust surface densities are denoted $\Sigma_{g,j}$ and $\Sigma_{d,j}$, respectively, where $j = 1, 2$ refers to the first and second bin as defined above. The dust surface density $\Sigma_{d,j}$ is a sum of contributions from both micron-sized dust and pebbles, with surface densities $\Sigma_{\mu,j}$ and $\Sigma_{pb,j}$ respectively. Accordingly, the total metallicity is defined by $\mathbb{Z}_j = \Sigma_{d,j}/\Sigma_{g,j} = (\Sigma_{\mu,j}+\Sigma_{pb,j})/\Sigma_{g,j}$, and the ``micron-scale" metallicity $\mathbb{Z}_{\mu,j} = \Sigma_{\mu,j}/\Sigma_{g,j}$. In the first bin, the water vapor mass fraction (of gas; $\Sigma_{vap}/\Sigma_{g,1}$) is given by $f_{vap}$. In the second, the ice mass fraction (of dust; $\Sigma_{ice}/\Sigma_{d,2}$) is given by $f_{ice}$, and assumed to be equal for both dust populations (this impacts our results insignificantly). Accordingly, the ice-to-rock ratio $R_{ice} = f_{ice}/(1-f_{ice})$. The first bin is continuously fed by water vapor and micron-sized dust from the inner disk, with constant vapor mass fraction $f_{vap,in}$ and metallicity $\mathbb{Z}_{\mu,in}$. Similarly, the second bin is fed by the inward drift of icy pebbles from the outer disk, with constant metallicity $\mathbb{Z}_{pb,out}$ and ice mass fraction $f_{ice,out}$. Below, we derive self-consistent equations for the evolution of $\Sigma_{\mu,j}$, $\Sigma_{d,j}$, $f_{vap}$, and $f_{ice}$ in time. The evolution of these quantities result from radial dust (\textbf{Section 2.3.1}) and vapor (\textbf{Section 2.3.2}) transport into, and between, the bins. A schematic of our model is provided in \textbf{Fig. 3}.

\subsubsection{Dust Advection, Drift, \& Growth}
\indent Particle size $a$ is directly related to the dimensionless Stokes number $St$, which quantifies the degree of dust-gas coupling. This coupling, in turn, determines the rate of radial dust transport, reflecting (in decretion disks) the competition between outward advection and inward drift due to azimuthal headwind drag. Assuming \textit{a priori} that particles of interest to us are in the Epstein drag regime \textcolor{blue}{(Armitage, 2020)}, we have 
\begin{equation}
St(a, r) = \sqrt{\frac{\pi}{8}}\frac{a\rho_0\Omega_k}{\rho_gc_s},
\end{equation}
where $\rho_0$ represents the material density of the particle (\textit{i.e.,} $\rho_{0,rock}\simeq 2000$ kg/m$^3$ for silicate dust; $\rho_{0,ice}\simeq 1000$kg/m$^3$ for water-ice), and $\rho_g = \Sigma_g/\sqrt{2\pi}h$ the volumic gas density. The particle radial velocity $v_{r,d}$ is given by the well-known expression \textcolor{blue}{(Nakagawa et al., 1986)}
\begin{equation}
v_{r,d} = \frac{v_{r,g} - |\epsilon|St (c_s^2/v_k)}{St^2 + 1}
\end{equation}
where $\epsilon$ represents the power law index of the disk pressure profile ($P = \Sigma c_s^2/\sqrt{2\pi}h\sim r^{-\epsilon}$). Given our parametrization of $\Sigma_g(r)$ and $T(r)$ in \textbf{Section 2.1}, $\epsilon$ evaluates to $3$. Note that the ($-$) sign in the second term above conveys the opposing radial directions of gas flow and particle drift. \\
\indent As mentioned above, we partition the dust budget between micron-sized particles and pebbles. For the former, we take $a_{\mu}\sim 10^{-6}$m in both bins. This translates to $St(a_{\mu,1})\sim St(a_{\mu,2})\sim 10^{-7}$, such that  $v_{r,d}\simeq v_{r,g}$.  Note that in the second bin (\textit{i.e.,} beyond the ice-line),  $St(a)$ depends on $\rho_0$ and thus $f_{ice}$ as $\rho_0 \simeq f_{ice}\rho_{0,ice} + (1-f_{ice})\rho_{rock}$. For simplicity, and given the largely illustrative nature of our work, we adopt characteristic pebble sizes of  $a_{pb,1} = 1$mm and $a_{pb,2} = 10$mm for the two bins. This is broadly consistent with results from \textcolor{blue}{Batygin \& Morbidelli (2020)}, suggesting most of the solid mass in circumplanetary decretion disk will be held in mm-scale pebbles, for which $v_{r,d}\sim0$ owing to balance between advection and drift. In the first bin (\textit{i.e.,} within the ice-line), $St(a_{pb,1}) \sim 10^{-4}$. In the second (\textit{i.e.,} beyond the ice-line), assuming an initial ice mass fraction of $f_{ice}=0.2$, we have $St(a_{pb,2})\sim 10^{-3}$. These $St$ values yield $v_{r,d}\sim 0.1$ and $\sim 1$m/s for the rocky and icy pebbles, respectively. \\
\indent In addition to dust radial transport, we account for particle growth from the micron-scale to pebbles within each bin, constituting a mass transfer between the two dust populations. Assuming monodisperse coagulation, the growth rate of a dust particle $\dot{a}$ is given by \textcolor{blue}{(Kornet et al., 2001; Brauer et al., 2008b)}
\begin{equation}
\dot{a} = \frac{\mathbb{Z}\rho_g h}{h_d\rho_0 }\Delta v,
\end{equation}
where the dust scale height $h_d$ is expressed in \textbf{Eq. 5}, $\mathbb{Z}$ represents the metallicity, and $\Delta v$ the velocity dispersion of dust particles. We assume dust relative velocities arise from turbulence, for which $\Delta v = \sqrt{3\alpha St}c_s$ \textcolor{blue}{(Ormel \& Cuzzi, 2007)}. Along with the expression for $St(a)$ in the Epstein regime (\textbf{Eq. 15}), the growth timescale $\tau_{grow} = a/\dot{a}$  in bin $j$ takes the form
\begin{equation}
\tau_{grow,j} \sim \frac{1}{\mathbb{Z}_j\Omega_k}ln\left(\frac{St(a_{pb,j})}{St(a_{\mu})}\right).
\end{equation}
The transfer in surface density between the two dust populations then, is given by $\dot{\Sigma}_{grow,j} = \Sigma_{\mu,j}/\tau_{grow}$. Note that, while the use of the micron-scale metallicity $\mathbb{Z}_\mu$ above would be consistent with our assumption of two dust populations, using the total metallicity $\mathbb{Z}$ instead likely lends itself to a more realistic estimate the pebble growth rate. 

\subsubsection{Vapor Advection \& Diffusion}
\indent Ice accumulation in bin 2 accompanies water vapor accumulation in bin 1. Vapor radial transport occurs by both advection at velocity $v_{r,g}$ (\textbf{Eq. 11}), driven by the turbulent viscosity $\nu$, as well as diffusion driven by turbulent mixing, at some characteristic velocity $v_{r,diff} \sim D/r$. Here, $D$ represents the turbulent diffusivity, which can be parametrized in terms of the turbulent viscosity $\nu$ via the turbulent Schmidt number $Sc$ as \textcolor{blue}{(\textit{e.g.,} Armitage, 2020)}
\begin{equation}
D = \nu/Sc.
\end{equation}
The Schmidt number quantifies the relative strength of angular momentum to mass transport by turbulence, and a range spanning from $\sim0.1$ to $10$ has been found collectively by various studies \textcolor{blue}{(Owen, 2014 \& references therein)}. Here, we assign a fiducial $Sc=2$, recognizing that, like $\alpha$, it remains a poorly determined quantity. The rate of diffusive vapor transport takes the form 
\begin{equation}
\dot{M}_{vap,D} =  \frac{2\pi r  \nu\Sigma_g}{Sc} \left(\frac{df_{vap}}{dr}\right) = \frac{2\pi r  \nu\Sigma_g}{Sc} \left(\frac{\Delta f_{vap}}{\Delta r}\right),
\end{equation}
where $\Delta f_{vap}$ represents the difference in the vapor mass fraction between bin 1 and either the inner disk ($< r_i - \Delta r$) or bin 2, and $\Delta r=0.01 R_{Hill}$ as mentioned above. For comparison, the rate of advective vapor transport is given by
\begin{equation}
\dot{M}_{vap,\nu} =  2\pi r \Sigma_g v_{r,g}f_{vap}.
\end{equation}
\begin{figure*} 
\centering
\scalebox{2}{\includegraphics{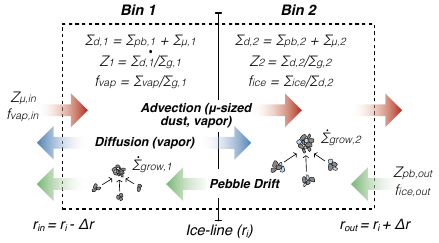}}
\caption{\textbf{Schematic of ``two-bin" model developed to simulate dust evolution (\textit{i.e.,} ice buildup) at the decretion disk ice-line,} including key model parameters. Dust and ice transport is driven by advection with the surrounding, decreting gas, and drift resulting from headwind drag. Vapor transport occurs by advection and diffusion. See \textbf{Section 2.3} for an in-depth description.}
\label{fig:Figure 3}
\end{figure*}
\subsubsection{Dust Evolution Equations}
\indent Here, we derive equations for $\dot{\Sigma}_{\mu,j}$, $\dot{\Sigma}_{pb,j}$, $\dot{f}_{vap}$, and $\dot{f}_{ice}$, self-consistently accounting for rock and water transport between disk annuli as well as pebble growth within both bins. We denote the centers of the bins by $r_1 = r_i - \Delta r/2$ and $r_2 = r_i + \Delta r/2$. Similarly, we denote the bin edges $r_{in} = r_i-\Delta r$ and $r_{out} = r_i+\Delta r$. Subscripts of all disk parameters below denote the place of evaluation. Thus, $\Sigma_{g,1} = \Sigma_g(r_1)$, $v_{r,g,in} = v_{r,g}(r_{in})$, and so forth. Recall that the area of both bins is given by $A\simeq 2\pi r_i\Delta r$. All dust transport terms below take the form of $\dot{M}/A$, where $\dot{M}$ takes the general form of $2\pi r v_r\Sigma$. Furthermore, while $v_{r,d}<0$ (pebbles drift inward), we formulate the equations below with $v_{r,d} = |v_{r,d}|$. That is, terms are specified as ``sources" or ``sinks" by the sign preceding them. \\
\indent Starting with the first bin, the evolution of $\Sigma_{\mu,1}$ is given by
\begin{equation}
\begin{split}
\dot{\Sigma}_{\mu, 1} &\simeq -\dot{\Sigma}_{grow,1} - \frac{r_1v_{r,g,1}\Sigma_{\mu,1}}{r_i\Delta r}  \\&+ \frac{r_{in} v_{r,g,in}\Sigma_{g,in}\mathbb{Z}_{\mu,in}}{r_i\Delta r},
\end{split}
\end{equation}
where the terms on the RHS represent, in order from left to right, pebble growth, advection out into the second bin, and introduction of new dust from the inner disk. The evolution of $\Sigma_{pb,1}$ is expressed as
\begin{equation}
\begin{split}
\dot{\Sigma}_{pb,1} &\simeq \dot{\Sigma}_{grow,1} - \frac{r_1\Sigma_{pb,1}v_{r,d,1}}{r_i \Delta r} \\&+ \frac{r_2\Sigma_{pb,2}v_{r,d,2}(1-f_{ice})}{r_i\Delta r},
\end{split}
\end{equation}
where the terms represent pebble growth, pebble drift into the inner disk, and introduction of pebbles drifting in from the second bin. The term $(1-f_{ice})$ reflects the sublimation of ice upon crossing the ice-line (we assume all rocky particles released are mm-scale pebbles). Finally, the evolution of $\Sigma_{vap}$ is given by
\begin{equation}
\begin{split}
\dot{\Sigma}_{vap} &\simeq -\frac{r_1\Sigma_{g,1}v_{r,g,1}f_{vap}}{ r_i \Delta r} + \frac{r_2\Sigma_{pb,2} v_{r,d,2}f_{ice}}{r_i\Delta r} \\&+ \frac{r_{in}\Sigma_{g,in}v_{r,g,in}f_{vap,in}}{ r_i \Delta r} \\&- \frac{r_1\nu_1\Sigma_{g,1}f_{vap}(2-f_{vap,in}/f_{vap})}{r_i\Delta r^2 Sc}
\end{split}
\end{equation}
where the terms represent the advection into the second bin, ice sublimation off of pebbles drifting in from the second bin, introduction of new vapor from the inner disk, and diffusion of vapor into the inner disk and second bin (recall $Sc =2$). The evolution of the vapor mass fraction $\dot{f}_{vap} = \dot{\Sigma}_{vap}/\Sigma_{g,1}$. \\
\indent Moving on to the second bin, we have 
\begin{equation}
\begin{split}
\dot{\Sigma}_{\mu, 2} &\simeq -\dot{\Sigma}_{grow,2} + \frac{r_1 v_{r,g,1}\Sigma_{\mu,1}}{r_i\Delta r} \\&- \frac{r_2v_{r,g,2}\Sigma_{\mu,2}}{r_i\Delta r} + \frac{0.5r_1v_{r,g,1}\Sigma_{g,1}f_{vap}}{ r_i \Delta r}\\& 
+\frac{0.5r_1\nu_1\Sigma_{g,1}f_{vap}}{r_i\Delta r^2 Sc},
\end{split}
\end{equation}
where the terms represent pebble growth, introduction of dust from the first bin, advection out to the outer disk, condensation of water vapor advected in from the first bin, and condensation of water vapor diffused in from the first bin. Regarding the latter two terms, $50\%$ of the introduced vapor is assumed to condense into micron-sized ice particles. For $\Sigma_{pb,2}$, we have
\begin{equation}
\begin{split}
\dot{\Sigma}_{pb,2} &\simeq \dot{\Sigma}_{grow,2} - \frac{r_2\Sigma_{pb,2}v_{r,d,2}}{r_i\Delta r} \\&+ \frac{0.5r_1v_{r,g,1}\Sigma_{g,1}f_{vap}}{ r_i \Delta r} + \frac{0.5r_1\nu_1\Sigma_{g,1}f_{vap}}{r_i\Delta^2 r Sc}\\&+ \frac{r_{out}\Sigma_{g,out}v_{r,d,r_{out}}\mathbb{Z}_{pb,out}}{r_i\Delta r},
\end{split}
\end{equation}
where the terms represent pebble growth, pebble drift into the first bin, condensation of water vapor advected and diffused from the first bin ($50\%$ onto existing particles), and introduction of new icy pebbles from the outer disk. Finally, for $\Sigma_{ice}$ we have
\begin{equation}
\begin{split}
\dot{\Sigma}_{ice} &\simeq - \frac{r_2\Sigma_{pb,2}v_{r,d,2}f_{ice}}{r_i\Delta r} - \frac{r_2v_{r,g,2}\Sigma_{\mu,2}f_{ice}}{r_i\Delta r} \\&+ \frac{r_1v_{r,g,1}\Sigma_{g,1}f_{vap}}{r_i \Delta r} + \frac{r_1\nu_1\Sigma_{g,1}f_{vap}}{r_i\Delta^2 r Sc} \\&+ \frac{r_{out}\Sigma_{g,r_{out}}v_{r,d,out}f_{ice,out}\mathbb{Z}_{pb,out}}{r_i\Delta r},
\end{split}
\end{equation}
where the terms represent pebble drift into the first bin, advection out to the outer disk, condensation of water vapor advected and diffused from the first bin, and introduction of new icy pebbles from the outer disk. The evolution of the ice mass fraction $\dot{f}_{ice} = \dot{\Sigma}_{ice}/\dot{\Sigma}_{d,2} = \dot{\Sigma}_{ice}/(\dot{\Sigma}_{\mu,2} + \dot{\Sigma}_{pb,2})$.\\
\indent At each timestep, all dust surface densities (\textit{i.e.,} metallicities) in both bins, as well as $\Sigma_{vap}$ (\textit{i.e.,} $f_{vap}$) and $\Sigma_{ice}$ (\textit{i.e.,} $f_{ice}$) are updated. Recall that $f_{ice}$ enters into the calculation of $St(a_{\mu,2})$ and $St(a_{pb,2})$ (\textbf{Eq. 15)} through $\rho_0$, and the metallicities $\mathbb{Z}_j$ into the calculation of $\tau_{grow,j}$ (\textbf{Eq. 18}). 

\subsection{Simulation Setup}
\indent Equipped with these coupled differential equations, we simulate the evolution of dust and water at the ice-line over a period of $t\simeq 10^4$ years, using a timestep $\Delta t\simeq 0.05$ year. The simulation is initialized assuming a disk with a flat metallicity profile set to $\mathbb{Z} = 0.01$ (\textit{i.e.,} the ISM value). The solid budget is assumed to be held primarily in pebbles ($95\%)$ as opposed to micron-sized particles ($5\%)$. Accordingly, the initial pebble metallicity $\mathbb{Z}_{pb,j}(t=0) = 0.95\times 0.01 = 0.95\%$, and the initial micron-sized metallicity $\mathbb{Z}_{\mu,j}(t=0) = 0.05\%$.  Similarly, at the bin boundaries $r_{in}$ and $r_{out}$ where vapor and ice are sourced, respectively, $\mathbb{Z}_{\mu,in} = 0.05\%$ and $\mathbb{Z}_{pb,out} = 0.95\%$. While the dust densities and thus metallicities in both bins change throughout the simulation, $\mathbb{Z}_{\mu,in}$ and $\mathbb{Z}_{pb,out}$ are kept constant. \\ 
\indent As for water vapor and ice fractions, $f_{vap,in} = f_{vap}(t=0)$ and $f_{ice,out} = f_{ice}(t=0)$, remaining constant across the simulation. In words, the vapor fraction of gas (ice fraction of pebbles) introduced into bin 1 (bin 2) throughout the simulation is equivalent to the initial vapor (ice) fraction therein. The former takes on a fiducial value of $0.001$, and the latter $0.2$, corresponding to an ice-to-rock ratio of $0.25$. These water-poor starting points are broadly consistent with our expectation from \textbf{Section 1.2} that the dust subsumed into the Jovian or Saturnian circumplanetary disks are expected to be subsolar in water content. 

\begin{figure*} 
\centering
\scalebox{1.1}{\includegraphics{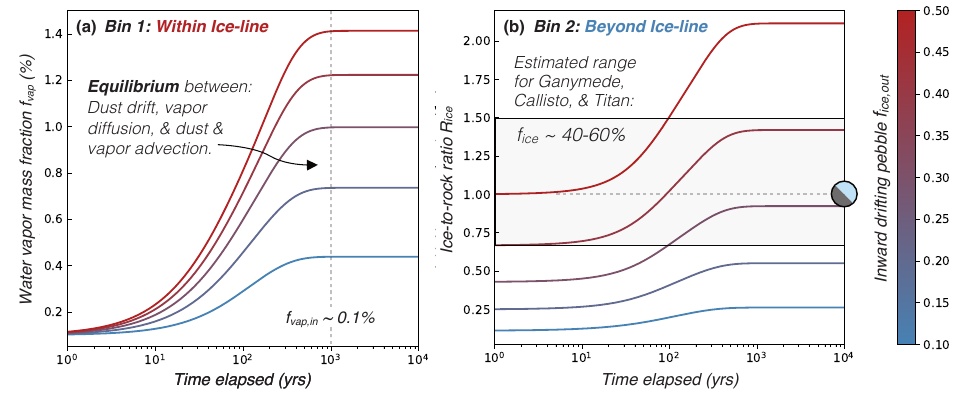}}
\caption{\textbf{Evolution of the (a) water vapor mass fraction $f_{vap}$ interior to the ice-line, and (b) ice-to-rock ratio $R_{ice}$ beyond it,} for different ice mass fractions of inward drifting icy pebbles from the outer disk $f_{ice,out}$. As is apparent, both $f_{vap}$ and $R_{ice}$ grow rapidly on a timescale of a few hundred years, reaching their steady-state values after $\sim 10^3$ years. Moreover, a range of $f_{ice,out}$ roughly between $\sim 0.25$ and $0.4$ ($R_{ice,out} = f_{ice,out}/(1-f_{ice,out})$ between $\sim 0.1$ and unity) yields final $R_{ice}$ values concordant with those estimated for the water-rich giant planet satellites. See text in \textbf{Section 3.1}.}
\label{fig:Figure 4}
\end{figure*}

\section{Results}
\indent Here, we show that an ice-to-rock ratio $R_{ice} = f_{ice}/(1-f_{ice})$ around or exceeding unity can be achieved at the ice-line within a timescale on the order of few times $\sim10^3$ years\textemdash less than the typical timescales associated with models for the accretion of the Galilean moons and  assembly of the Laplace resonance by disk-driven migration \textcolor{blue}{(\textit{e.g.,} Peale \& Lee, 2002; Sasaki et al., 2010; Shibaike et al., 2019; Batygin \& Morbidelli, 2020; Yap \& Batygin, 2025)}. This is done while maintaining realistic values for the vapor mass fraction $f_{vap}$ ($\lesssim 1$ to $2\%$). In \textbf{Section 3.1}, we explore the impact of the $f_{vap,in}$ and $f_{ice,out}$ (\textit{i.e.,} the source terms for water; see \textbf{Fig. 3}) on the evolution of $f_{vap}$ and $R_{ice}$. We proceed to consider the evolution of metallicities in \textbf{Section 3.2}, before observing the dependence of our results on the uncertain quantities pertaining to turbulence, namely $Sc$ and $\alpha$, in \textbf{Section 3.3} and \textbf{3.4}, respectively.
\subsection{Water Ice \& Vapor Fractions}
\indent Water delivery to the ice-line is facilitated by both the advection of new vapor from the inner disk, at mass fraction (of gas) $f_{vap,in}$, and inward drift of new ice from the outer disk, at mass fraction (of pebbles) $f_{ice,out}$. Recall that $f_{vap,in} = f_{vap}(t=0)$ and $f_{ice,out} = f_{ice}(t=0)$, remaining constant across the simulation. Here, we present simulation results across a range of these two quantities.\\
\indent Starting with the latter, we display in \textbf{Fig. 4} the evolution of $f_{vap}$ (within the ice-line; \textbf{4a}) and $R_{ice}$ (beyond the ice-line; \textbf{4b}) as a function of time, for $f_{ice,out}$ between $0.1$ and $0.5$, and $f_{vap,in} = 0.1\%$. This range of $f_{ice,out}$ corresponds to $R_{ice,out} = f_{ice,out}/(1-f_{ice,out})$ between $\sim 0.1$ and unity (equivalent to the initial $R_{ice}(t=0)$ in bin 2). \\
\indent As is evident, both $f_{vap}$ and $R_{ice}$ grow rapidly on a timescale of a few hundred years, reaching their equilibrium values in only $\sim1$ kyr, an order of magnitude less than the characteristic timescale for disk evolution $\tau_\Sigma\sim 10^4$ (\textbf{Eq. 14}; see \textbf{Section 2.1}). Thus, the assumption of a steady-state disk [\textit{i.e.,} constant $\Sigma_g(r)$ and $T(r)$] is valid for our work. As discussed in \textbf{Section 2.2}, ice and vapor accumulation occur in tandem, as water that makes its way to the ice-line goes back and forth across it (\textbf{Fig. 2}). That said, key to our simulation reaching steady-state is vapor diffusion, in particular that towards the inner disk. Without it, the only sink for water accumulated at the ice-line would be the advection of micron-sized icy particles towards the outer disk, far insufficient to prevent $f_{vap}$ and $R_{ice}$ from reaching un-physical values.\\
\indent Before moving on, we emphasize that the steady state established is an outcome of our simple two-bin model, which does not account for mass conservation\textemdash water and dust removed from the system are assumed to be wholly lost. In reality, water is expected to continually accumulate at the ice-line, perhaps leading to an extended water-rich disk region. While useful to gauge the timescale on which water accumulates at the ice-line, as well as the interplay of disk parameters, our model ultimately does not capture the complexity afforded by a global model.\\
\indent A clear observation from \textbf{Fig. 4} is that larger $f_{ice,out}$ leads to larger equilibrium values for $f_{vap}$ and $R_{ice}$. This simply reflects that, with a larger source term for water, a higher $f_{vap}$ (and thus $R_{ice})$ is required for vapor diffusion to stabilize the system. Starting at the fiducial initial value of $0.1\%$, the vapor fraction $f_{vap}$ rises to $\gtrsim 0.4\%$, exceeding $1\%$ for $f_{ice,out}$ of $\gtrsim0.3$. As for $R_{ice}$, it appears the final $R_{ice}$ reached is $\sim$ twice that of $R_{ice}(t=0)$. That is, the ice-line (given our aforementioned choices of disk parameters) has the capacity to double the initial ice-to-rock ratio of dust in the disk. Ganymede, Callisto, and Titan are estimated to have $f_{ice}\sim 40$ to $60\%$ (\textit{i.e.,} $R_{ice}\sim 0.7$ to $1.5$). From \textbf{Fig. 4b}, it is clear that satellitesimals serving as their building blocks can be characterized by this range of $R_{ice}$ for $f_{ice,out}$ between $\sim0.25$ and $0.4$, corresponding to $R_{ice,out} = R_{ice}(t=0)$ from $\sim 0.33$ to $0.67$. A disk with solids possessing, as a whole, an ice-to-rock ratio $\sim 0.5$ can beget a satellite formation region with a composition akin to that observed in the icy satellites. \\
\indent Moving on to $f_{vap,in}$, we showcase in \textbf{Fig. 5} the final $R_{ice}$ reached as a function of $f_{vap,in}$ between $0.1\%$ and $1\%$. This is done for the range of $R_{ice,out}$ explored in \textbf{Fig. 4}. The key observation here is that while higher$f_{vap,in}$ leads to larger final $R_{ice}$, it is secondary to $f_{ice,out}$. Indeed, increasing $f_{vap,in}$ by an order of magnitude ($0.1\%$ to $1\%$) elevates the final $R_{ice}$ by a factor of $\lesssim 1.5$, while increasing $f_{ice,out}$ by a factor of $\sim 2$ leads to an increase in the final $R_{ice}$ by a factor of $\sim2$ to $3$ (\textbf{Fig. 4b}). This indicates that the primary source of water delivery to the ice-line is the drift of icy pebbles from the outer disk, not the advection of water vapor from the inner disk (at least given our simulation setup). \\
\indent All in all, \textbf{Fig. 5} demonstrates that in $<10^4$ years, $R_{ice}$ can increase by a factor of $\gtrsim 2$ from its initial value for typical disk parameters, specifically our choice of $Sc=2$ and $\alpha=10^{-4}$ (see \textbf{Sections 3.3 \& 3.4} below).

\begin{figure} 
\scalebox{0.95}{\includegraphics{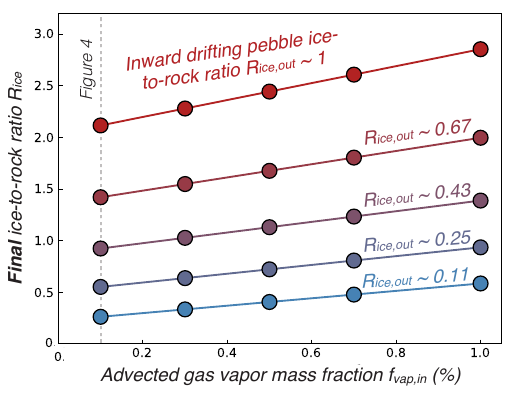}}
\caption{\textbf{The final (\textit{i.e.,} steady-state) ice-to-rock ratio $R_{ice}$ as a function of the mass fraction of advected water vapor from the inner disk $f_{vap,in}$, for various $R_{ice,out}$ of icy pebbles drifting inward for the outer disk.} The vertical dashed line at $f_{vap,in}$ corresponds to the simulation results displayed in \textbf{Fig. 4}. Clearly, the primary control on the final $R_{ice}$ of the dust reservoir beyond the ice-line is $R_{ice,out}$, not $f_{vap,in}$, indicating most of the water delivered comes from the outer disk in the form of pebbles. See text in \textbf{Section 3.1}.}
\label{fig:Figure 5}
\end{figure}

\subsection{Total Mass Accumulated}
\indent In addition to the ice-to-rock ratio of the dust/solid reservoir, it is worthwhile to consider how its total mass at steady-state. In \textbf{Fig. 6}, we display the evolution of the total, pebble, and micron-sized metallicities beyond the ice-line (\textit{i.e.,} $\mathbb{Z}_2$, $\mathbb{Z}_{pb,2}$, and $\mathbb{Z}_{\mu,2}$, respectively) as a function of time for the case with $f_{ice,out} = 0.2$. Starting from the ISM value of $\sim1\%$, $\mathbb{Z}_2$ climbs just under $2\%$ at steady-state. Close to $3/4$ of the dust is held in pebbles, with the remaining $1/4$ hosted in micron-sized particles. The smaller micron-sized population lends credence to our assertion above that advection alone is insufficient as a sink for water.\\
\indent The steady-state mass held in pebbles by $\sim 10^3$ years is roughly coincident with $\sim10\%$ the mass of Ganymede, Callisto, and Titan (each $\sim 10^{23}$kg). The formation of satellitesimals (\textit{i.e.} satellite building blocks) is expected to occur by concentration and gravitational self-collapse of pebble clouds \textcolor{blue}{(\textit{e.g.,} Youdin \& Goodman, 2005)}. The rapid rate with which steady-state is reached indicates replenishment of $\mathbb{Z}_{pb}$ will outpace depletion by satellitesimal formation, suggesting it will take place amidst a $\mathbb{Z}_{pb}$ backdrop of $\sim 1.5\%$ given an annulus of width $\Delta r = 0.01R_{Hill}$. While satellitesimal formation is predicted to occur on a timescale of a few hundred to thousand orbits \textcolor{blue}{(\textit{e.g.,} Ostertag \& Flock, 2025)}, further growth by mutual collisions occurs on a timescale of order $\sim10^4$ years \textcolor{blue}{(Batygin \& Morbidelli, 2020)}. The latter is thus expected to be the ``bottleneck" in satellite formation. In the \textbf{Appendix}, we provide a derivation of the length-scale associated with satellitesimal formation, showing it to be less than $\Delta r$.\\

\indent Within the ice-line, almost all the solid mass is held in pebbles, with $\mathbb{Z}_1\simeq \mathbb{Z}_{pb,1} $ reaching $\sim 8.5\%$. This reflects both the shorter pebble growth timescale within the ice-line $\tau_{grow,1}$ [$St(a_{pb,1})\sim 0.1St(a_{pb,2})$; see \textbf{Section 2.3.1} \& \textbf{Eq. 18}], as well as our simplifying assumption that all rocky particles released by ice sublimation off of inward drifting pebbles are pebble-sized (\textit{i.e.,} $\sim 1$mm; see last term in \textbf{Eq. 23}). Note that $\tau_{grow,1}$ also decreases with $\mathbb{Z}_1$. The higher metallicity interior to the ice-line reflects the aforementioned ``traffic jam" effect in \textbf{Section 2.2}, expected to occur in accretion disks as well \textcolor{blue}{(Drażkowska \& Alibert, 2017)}. 
\begin{figure} 
\scalebox{1.15}{\includegraphics{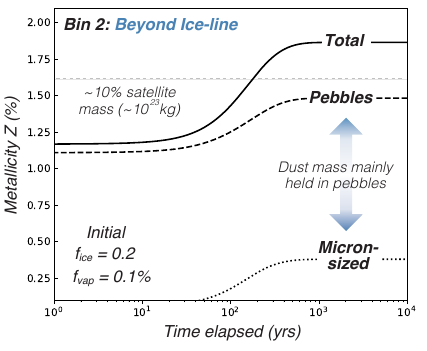}}
\caption{\textbf{Evolution of the total (solid), pebble (dashed), and micron-sized metallicities (dotted) metallicities}, for $f_{vap,in} = 0.1\%$ and $f_{ice,out}=0.2$. At steady-state, $\sim3/4$ of the total dust mass is held in pebbles. Moreover, a solid mass $>10\%$ that of Ganymede, Callisto, and Titan is accumulated in only a few hundred years within the annulus of width $\Delta r =0.01 R_{Hill}$. See text in \textbf{Section 3.2}.}
\label{fig:Figure 6}
\end{figure}
\subsection{Impact of the Turbulent Schmidt Number}
\indent The Schmidt number $Sc$ controls the strength of turbulent diffusivity $D$ (given a turbulent viscosity of $\nu$; \textit{i.e.,} $\alpha$) and is perhaps the most dubious quantity in our dust evolution model. As noted in \textbf{Section 2.3.2}, reported values for $Sc$ from the literature span $\sim$ two orders of magnitude, roughly centered on the fiducial value of $Sc=2$ adopted. Motivated by both its uncertainty, and the importance of vapor diffusion in our dust evolution model (see \textbf{Section 3.1} above), we explore here its impact on our simulation outcomes. \\
\indent Higher $Sc$ translates to lower diffusivity $D$. As such, we expect higher $Sc$ to yield larger values of $f_{vap}$ at steady-state, in the same way that lower turbulent viscosity $\nu$ calls for a larger surface density $\Sigma_g$ to produce a given mass flow rate. This is exactly what we observe in \textbf{Fig. 7a}, showcasing the evolution of $f_{vap}$ over time for increasing $Sc$. While the steady-state $f_{vap}$ lies below $\sim0.5\%$ for $Sc\lesssim 1$, it rises dramatically to $\gtrsim 3\%$ for $Sc = 10$. \textbf{Fig. 7b} displays the corresponding evolution of $R_{ice}$, and \textbf{Fig. 7c}, that of $\mathbb{Z}_2$ and $\mathbb{Z}_{pb,2}$. Overall, it appears that $Sc$ controls the partitioning of water between the vapor and solid phase (\textit{i.e.,} storage within or beyond the ice-line). In particular, lower $Sc$ leads to more water being stored as ice. \\
\indent Interestingly, closer inspection reveals that $R_{ice}$ at steady-state decreases with increasing $Sc$ up to a value of $\sim3$, but exhibits an opposite trend for $Sc\gtrsim3$. The same pattern is observed for the metallicities, and indicates the presence of two regimes in $D$.  For relatively low $Sc$ ($\lesssim 3$), increasing $Sc$ leads to less water vapor diffusion across the ice-line (\textit{i.e.,} more water stored as vapor), thereby diminishing $R_{ice}$, and by extension $\mathbb{Z}_2$ and $\mathbb{Z}_{pb,2}$.  Here, the transport of water across the ice-line with increased diffusivity (\textit{i.e.,} lower $Sc$) outweighs the accompanying loss of water by the system as a whole by diffusion into the inner disk (\textit{i.e.,} out of our simple two-bin model), leading to a negative correlation between $R_{ice}$ and $Sc$. In the high $Sc$, or low $D$, regime, we interpret the positive correlation between $R_{ice}$ and $Sc$ as reflecting the opposite: diffusive lost of vapor from the system outweighs the transport of water across the ice-line, and thus prevents a net rise in $R_{ice}$ from increasing $D$. That is, although a larger fraction of the vapor present in bin 1 is being  transported across the ice-line, there is less vapor therein at steady-state. This is supported by the drastic drop in the steady-state $f_{vap}$ between the case with $Sc = 10$ and $Sc \sim3.3$. Note, nonetheless, that the trend in the high $Sc$ regime appears weak (the absolute change in $D$ in going from $Sc\sim 3$ to 10 is small). More importantly, this trend is likely un-physical, reflecting our assumption that any vapor lost to the inner disk does not find its way back to the ice-line. In a global model wherein mass conservation across the disk can be implemented, the parameter $f_{vap,in}$ would not remain constant at its initial value. \\
\indent A key takeaway from \textbf{Figs. 7b \& 7c} is that for a given $f_{vap,in}$ and $f_{ice,out}$, lower $Sc$ can facilitate higher $R_{ice}$ and (less importantly) $\mathbb{Z}_2$. That is, stronger turbulent diffusion can compensate for smaller rates of water delivery to the ice-line (\textit{i.e.,} smaller $f_{vap,in}$ and $f_{ice,out}$; see \textbf{Section 3.1}). Decreasing $Sc$ by an order of magnitude from $\sim 1$ to $\sim 0.1$, keeping all other parameters constant, leads to an increase in the final $R_{ice}$ by a factor of $\sim 2$ (and $\mathbb{Z}_2$ by a factor of $\sim 1.5$). Thus, $f_{ice,out}$ on the order of $\sim 0.1$ or $0.2$, while seemingly too small to yield a dust reservoir with $R_{ice}\sim1$ (\textbf{Fig. 4b}), is only so for $Sc=2$. If a lower $Sc$ is applicable, an initial $R_{ice}(t=0)\lesssim 0.2$ may be sufficient to yield the water-rich compositions observed for the satellites today. Although not shown, the metallicity within the ice-line (\textit{i.e.,} $\mathbb{Z}_1$) plateaus at $\sim 8\%$ regardless of $Sc$, and is essentially held in pebbles ($\mathbb{Z}_1 \simeq \mathbb{Z}_{pb,2}$), as found in our fiducial simulation (see \textbf{Section 3.2}). \\
\indent As a final point, the timescale for steady-state to be established increases with larger $Sc$. This makes intuitive sense\textemdash as vapor diffusion is responsible for establishing steady-state (see \textbf{Section 3.1}), a larger diffusivity translates to a shorter time to reach it (\textit{i.e.,} the steady-state vapor fraction of gas in bin 1 is lower). Increasing $Sc$ by two orders of magnitude extends the said time by only a factor of a few.
\begin{figure} 
\scalebox{1.1}{\includegraphics{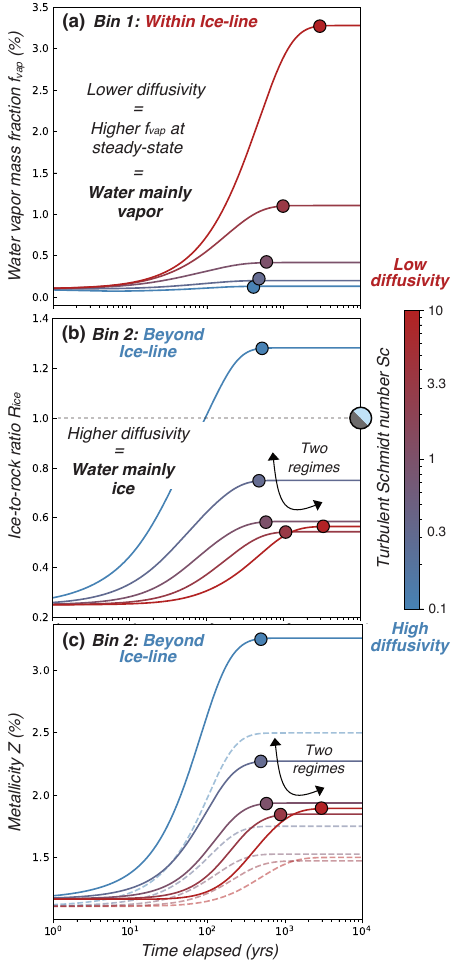}}
\caption{\textbf{Evolution of  (a) $f_{vap}$, (b) $R_{ice}$, and (c) $\mathbb{Z}_2$ (along with $\mathbb{Z}_{pb,2}\sim3\mathbb{Z}_2/4$; dashed) for turbulent Schmidt numbers $Sc$ ranging between $0.1$ and $10$.} As is evident, higher $Sc$ (lower diffusivity) yields higher steady-state $f_{vap}$. For $R_{ice}$ and $\mathbb{Z}_2$, two regimes are apparent: (i) for $Sc\lesssim3$, both quantities decrease with increasing $Sc$; (ii) for $Sc\gtrsim3$, the opposite trend is true. See text in \textbf{Section 3.4}.}
\label{fig:Figure 7}
\end{figure}
\subsection{Impact of the $\alpha$ Turbulence Parameter}
\indent The Shakura-Sunyaev $\alpha$ turbulence parameter, while immensely useful, is notoriously uncertain. Like $Sc$ above, the typical range for $\alpha$, informed by both models of turbulence generation in protoplanetary disks [\textit{e.g.,} via the magnetorotational or vertical shear instability (MRI or VSI); \textcolor{blue}{Balbus \& Hawley, 1991; Nelson et al., 2013}] and observations thereof by the Atacama Large Millimeter Array (ALMA; \textcolor{blue}{see recent review by Rosotti et al., 2023}), spans two orders of magnitude from $\sim10^{-4}$ to $10^{-2}$. Here, we explore its impact on our results, allowing it to range from $10^{-5}$ to $10^{-3}$ (\textit{i.e.,} centered on our fiducial value of $\alpha=10^{-4}$). For a given surface density $\Sigma_g$, variations in $\alpha$ translate to variations in the mass decretion/accretion rate $\dot{M}$. As such, variations in $\alpha$ can be broadly taken as a proxy for those in $\dot{M}$, which also constitutes a highly uncertain quantity. \\
\indent Given that $Sc\sim \nu/D$ and $\nu\sim \alpha^{4/3}$ (see \textbf{Eqs. 8 \& 9}), we expect an increase in $\alpha$ at constant $Sc$ to have a similar impact as a decrease in $Sc$ at constant $\nu$. Variations in $\alpha$ (or $\nu)$, nonetheless, introduce an additional layer of complexity\textemdash it directly governs the rate of outward gas advection $v_{r,g}\sim \nu/r$ (see \textbf{Eq. 11}), which in turn affects the rate of inward pebble drift (see \textbf{Eq. 16}). In particular, higher $\alpha$ leads to faster gas advection, and slower pebble drift. Note that at $\alpha\gtrsim$ few times $10^{-3}$ (not explored here), $v_{r,d}$ can be positive for pebble sizes at the mm-scale, such the differential equations in \textbf{Section 2.3.2} must be modified. \\
\indent In \textbf{Fig. 8}, we display the evolution of \textbf{(a)} $f_{vap}$, \textbf{(b)}, $R_{ice}$, \textbf{(c)} $\mathbb{Z}_1 (\mathbb{Z}_{pb,1})$, and \textbf{(d)} $\mathbb{Z}_2 (\mathbb{Z}_{pb,2})$ over time for the said range of $\alpha$. As expected, the key observations are broadly consistent with those from \textbf{Fig. 9}. Higher $\alpha$, translating here to \textit{both higher diffusivity and stronger advection}, leads to more water being stored as ice beyond the ice-line in steady-state. In \textbf{Fig. 8b}, we observe that an increase in $\alpha$ from $10^{-4}$ to $10^{-3}$ yields a rise in the final $R_{ice}$ by a factor of $\sim 1.5$, less than the factor of $\sim 2$ accompanying the decrease of $Sc$ by an order of magnitude (see \textbf{Section 3.3} \& \textbf{Fig. 7b}). This reflects the subtle difference between variations in $Sc$ and $\alpha$. In particular, while increasing the latter (like decreasing the former) leads to an increase in $D$, it also slows the inward drift of icy pebbles, which we have established is the dominant source for water delivery to the ice-line (see \textbf{Section 3.1}). Moreover, at low $\alpha$ (\textit{i.e.,} $\sim 10^{-5}$), the trend reversal observed in \textbf{Section 3.3} is absent (see \textbf{Figs. 8b \& 8d}). We interpret this as reflecting that increased advection (of vapor across the ice-line) buffers $R_{ice}$ and $\mathbb{Z}_2$ against increased diffusive vapor loss of the system as a whole in the low $D$ regime. \\
\indent Additional evidence for the different impact between $Sc$ and $\alpha$ is given by $\mathbb{Z}_1\simeq \mathbb{Z}_{pb,1}$ in \textbf{Fig. 8c}. As discussed above, $Sc$ alone has virtually no impact on the steady-state value of $\mathbb{Z}_1$. Clearly, this is not the case with $\alpha$, as higher $\alpha$ leads to higher $\mathbb{Z}_1$. This reflects the slowing of the pebble drift rate due to enhanced outward gas velocity $v_{r,g}$, leading to a higher solid budget at steady-state. Finally, we observe that lower $\alpha$ results in a longer time for steady-state to be established, as with higher $Sc$ in \textbf{Fig. 7}. This is in accord with the notion that the timescale for the system to settle is primarily controlled by the diffusivity $D$. Also, it appears different parameters (\textit{e.g.,} $f_{vap}$ and $\mathbb{Z}_j$) reach steady-state at slightly different times. This is not unexpected given their different sources and sinks at play.

\subsection{Caveats regarding Dust Transport}
\indent In our treatment of dust transport (see \textbf{Section 2.3.1}), we accounted for outward advection with the decreting gas, and inward drift due to azimuthal headwind drag from the pressure supported, and hence sub-Keplerian, gas flow. For simplicity, we had neglected two effects that, when coupled, impact the degree of dust retention at the ice-line, and thus the steady-state ice-to-rock ratio (\textit{i.e.,} the final $R_{ice}$) achieved. These are (i) the back-reaction of dust onto the gas \textcolor{blue}{(\textit{e.g.,} Nakagawa et al., 1986; Cuzzi et al.. 1993; Dipierro et al., 2018)}, and (ii) turbulent dust diffusion \textcolor{blue}{(Youdin \& Lithwick, 2007)}. \\
\indent Regarding (i), when metallicities significantly exceed the solar value of $\sim 1\%$, as observed within the ice-line (\textit{i.e.,} in bin 1) of our fiducial case ($\mathbb{Z}_1 \sim 8.5\%$; see \textbf{Section 3.2}), dust drag onto its surrounding gas leads to an increase in its orbital velocity, thus decreasing headwind drag and pebble inward drift velocities. This effect plays a crucial role in the development of dust-gas instabilities \textcolor{blue}{(\textit{e.g.,} Youdin \& Goodman, 2005; Squire \& Hopkins, 2018)} thought to result in planetesimal/satellitesimal formation (see \textbf{Appendix}). Importantly, the accumulation and settling/sedimentation of dust alone would aid its retention\textemdash within the dust layer, particles are shielded from headwind drag. Greater retention of rocky pebbles within the ice-line ultimately lowers the ice-to-rock ratio beyond it, resulting from turbulent diffusion of those pebbles back across the ice-line [effect (ii)], and enhanced advection of rocky micron-sized particles. The latter would take place due to pebble fragmentation (replenishing the micron-sized dust population), also neglected in this study. \\
\indent Future inclusion of these effects should take place within a global model for dust evolution in circumplanetary disks, accounting for mass conservation and disk evolution (\textit{e.g.,} the movement of the ice-line due to dust/opacity buildup). Such a global model would yield a more realistic picture of the spatial extent and water abundance of the water-rich solid reservoir.

\section{Discussion}
\subsection{Icy Satellite D/H Ratios}
\indent We have demonstrated that, for a reasonable set of disk parameters, the ice-line in a decreting circumplanetary disk can facilitate the emergence of an ice-rich solid reservoir just beyond it. This ice-rich composition is envisioned to be sampled by satellitesimals constituting the precursor objects to the icy giant planet satellites today, formed via the agglomeration and collapse of pebble clouds (see \textbf{Appendix}; \textcolor{blue}{Mosqueira \& Estrada, 2003; Batygin \& Morbidelli, 2020}). \\
\indent Formation models for icy satellites must account for their hydrogen isotopic compositions, namely their D/H ratios. Indeed, D/H ratios of planetary materials widely reflect those of their building blocks, and serve as broad tracers of provenance in the early SS \textcolor{blue}{(\textit{e.g.,} Geiss \& Reeves, 1981; Lécluse \& Robert, 1994; Drouart et al., 1999 Robert et al., 2000; Yang et al., 2013)}. In particular, a D/H ``gradient" is thought to have characterized the circumsolar disk, increasing with radial distance from the nascent Sun. The notion of such a gradient is underpinned by theoretical investigations of grain-surface and ion-molecule reactions at the low temperatures ($T<50$K) in dense molecular clouds, suggesting hydrogen isotopic fractionation attending those reactions result in highly deuterated water-ice, with D/H $\gtrsim 10^{-3}$ \textcolor{blue}{(\textit{e.g.,} Solomon \& Woolf, 1973; Brown \& Millar, 1989; Roberts et al., 2004)}. Following the onset of cloud collapse and circumsolar disk formation, rapid isotopic equilibration between sublimed water vapor and surrounding $H_2$ in the hot ($T\sim 1000K$) inner portions of the disk is envisioned to deplete the former in D. The gas-phase isotopic exchange is described by the reaction
\begin{equation}
H_2O + HD \rightleftharpoons HDO + H_2.
\end{equation}
Moving towards the outer disk, in particular the ice-line, D/H ratios are expected to rise, reflecting increasing contributions from pristine, high-D ices directly inherited from the SS parent molecular cloud. Notably, the inferred D/H gradient is supported (or at least consistent) with D/H ratios measured in carbonaceous chondrites, Jupiter-family comets, and Oort cloud comets ($\sim$ $1$ to  $3 \times 10^{-4}$; \textcolor{blue}{\textit{e.g.,} Bockeleé-Morvan et al., 1998, 2012, 2015; Altwegg et al., 2015; Clark et al., 2019}), close to an order of magnitude higher than that inferred for the bulk SS (\textit{i.e.,} $H_2$ gas), at $\sim 2\times 10^{-5}$ \textcolor{blue}{(Geiss \& Gloecker, 1998; Mahaffy et al., 1998; Lellouch et al., 2001)}.\\
\begin{figure*} 
\centering
\scalebox{1}{\includegraphics{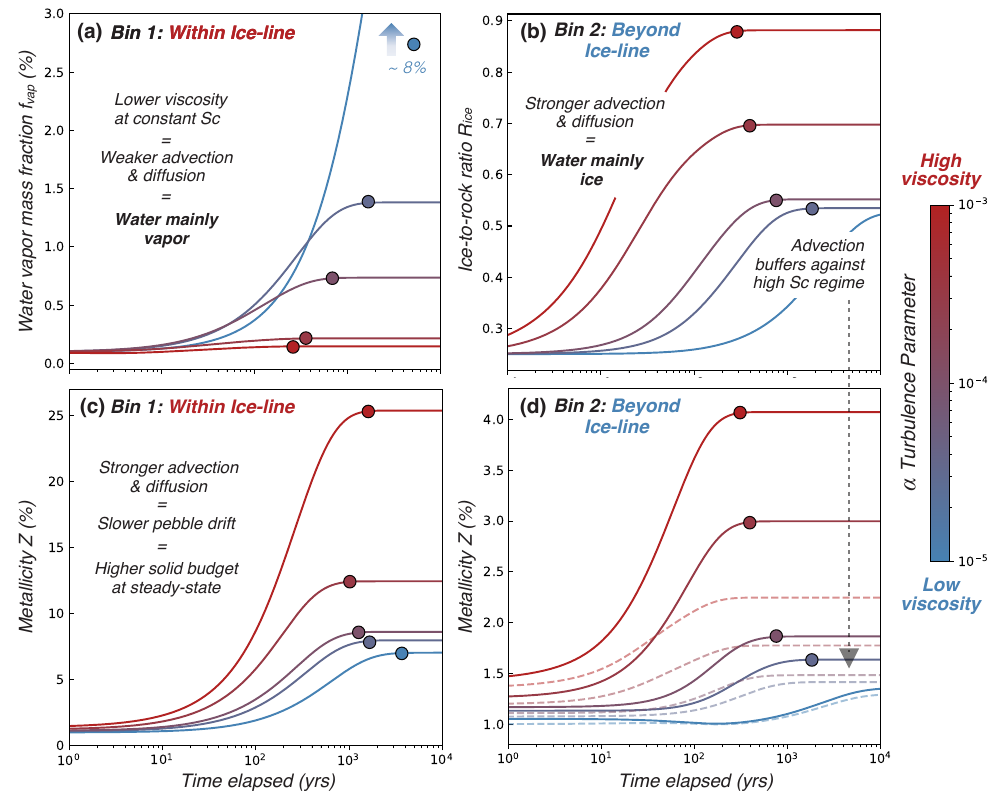}}
\caption{\textbf{Evolution of (a) $f_{vap}$, (b) $f_{ice}$. (c) $\mathbb{Z}_1\simeq \mathbb{Z}_{pb,1}$, and (d) $\mathbb{Z}_2$ and $\mathbb{Z}_{pb,2}$ for $\alpha$ turbulence parameters ranging between $10^{-5}$ and $10^{-3}$.} Evidently, lower $\alpha$ at constant Schmidt number $Sc$ (\textit{i.e.,} lower diffusivity) leads to more water being stored as vapor within the ice-line than ice beyond it, and vice versa. Due to stronger gas advection and thus slower pebble drift velocities for higher $\alpha$, steady-state metallicities in both bins are larger. See text in \textbf{Section 3.4}.}
\label{fig:Figure 8}
\end{figure*}
\indent Measured D/H ratios of giant planet satellites are virtually confined to those of Saturn, and attributed to both the \textit{Cassini} spacecraft \textcolor{blue}{(Waite et al., 2009; Clark et al., 2019)} and more recently, the James Webb Space Telescope \textcolor{blue}{(Brown et al., 2025)}. With the exception of Titan, these ratios reflect those of surface water-ice, and range between $\sim 2$ to $4 \times 10^{-4}$. Under the assumption that surface ice D/H has not been significantly modified by space weathering, and as such, is representative of the bulk interior, these ratios suggests Saturnian satellite building blocks were D-enriched, comprising water-ice that escaped high-temperature isotopic equilibration. For Titan, D/H ratios have been measured in atmospheric $CH_4$ and $C_2H_2$ (acetylene), sitting at $\sim 1.6\times 10^{-4}$ \textcolor{blue}{(Nixon et al., 2012)} and $\sim 1.2\times 10^{-4}$ \textcolor{blue}{(Bézard et al., 2024)}, respectively. Although complicated by the details of its surface-atmospheric evolution (\textit{i.e.,} whether or not the D/H of these species reflect that of its interior water), Titan likely bears a D/H signature of water in broad agreement with those of its neighbors, similarly hinting at a substantial contribution of pristine ices to its accretion. \\
\indent While hydrogen isotopic compositions of the Galilean satellites are lacking, recent re-analysis of data from the Near-Infrared Mapping Spectrometer (NIMS) on the \textit{Galileo} spacecraft has constrained the D/H ratio of Callisto's surface ices to $\gtrsim10^{-4}$ \textcolor{blue}{(Clark et al., 2019)}. Accordingly, all inferences of giant planet satellite D/H ratios made thus far paint a picture of highly deuterated formation regions, consisting of water-ice that have evaded high temperatures since its time in frigid molecular cloud. \\
\indent The elevated D abundance inferred for the giant planet satellites begs the question as to whether the ice-rich reservoir proposed to lie beyond the circumplanetary disk ice-line can be D-enriched. Answering this question requires consideration of the journey taken by the accumulated water from the circumsolar disk. Recall that this water is attributed to both (i) water vapor from the inner disk, derived from the sublimation of icy pebbles introduced by meridional flows onto the giant planet, and (ii) inward drifting icy pebbles from the outer disk. Icy pebbles in the circumsolar disk likely bear a D-enriched signature akin to carbonaceous chondrites (and comets), with D/H $\sim 10^{-4}$. Although source (i), the water vapor, is derived from these pebbles, it has passed high temperatures ($\gtrsim$ a few $100K$; see \textbf{Fig. 1}) close in to the giant planet. This may have led to its isotopic equilibration with the surrounding $H_2$ gas, resulting in a low (\textit{i.e.,} $\sim 10^{-5}$) D/H signature. Nonetheless, the extent of equilibration depends on the yet unconstrained pressures and temperatures attending material infall onto the circumplanetary disk, setting the kinetics for isotopic exchange (\textbf{Eq. 28}; see also \textbf{Section 4.2} below). Moving on to source (ii), if inward drifting icy pebbles were directly derived from the circumsolar disk through, say, capture at its outer edge by drag \textcolor{blue}{(Homma et al., 2020)}, they would have retained their high D/H ratios. While the water introduced into the ice-rich reservoir by these pebbles is envisioned to pass back and forth across the ice-line (see \textbf{Fig. 2} \& \textbf{Section 2.2}), the isotopic equilibration between gas-phase water and $H_2$ at temperatures close to $\sim 170K$ is likely quenched. \\
\indent In our model, as discussed in \textbf{Section 3.1}, icy pebbles from the outer circumplanetary disk constitute the primary source of water buildup at the ice-line. Accordingly, the proposed scenario for the formation of icy satellites accommodates for their inferred high-D compositions. Future constraints on satellite D/H ratios by, for instance, the JUpiter ICy moons Explorer (JUICE; \textcolor{blue}{Grasset et al., 2013}), will refine models of water delivery to satellite formation regions.
\subsection{CO-CH$_4$ Conversion as a Source of Water-Ice}
\indent While our proposed ``cold finger" mechanism can readily account for the high ice-to-rock ratio inferred for the three giant planet satellites, it is worthwhile to consider if CO-CH$_4$ conversion can contribute to water production. That is, if there is an astrophysical setting in which \textbf{Eq. 1} is thermodynamically favored to proceed to the right, \textit{and} is not kinetically inhibited to do so. The satisfaction of the latter condition depends on the existence of sufficient collisional energy to overcome an activation barrier, which value reflects the electronic energies of the intermediate molecules/states that provide the most favorable reaction path.  Although the kinetics (\textit{i.e.,} the intermediate states constituting the reaction ``bottleneck") are far less understood than the thermodynamics \textcolor{blue}{(Moses, 2014; Wang et al., 2016)}, a ``quench temperature" of $T_Q\sim 1000K$ below which CO-CH$_4$ conversion is effectively halted is consistent with the observed CO abundance in the atmosphere of Jupiter \textcolor{blue}{(Cavalie et al., 2023)}. As we will show, \textit{pressure is the true roadblock to efficient CO-CH$_4$ conversion,} which generally requires $P>0.1$ bar (see \textbf{Eq. 2} \& \textbf{Fig. 9}). \\
\indent The high temperatures and pressures required for efficient CO-CH$_4$ conversion appear incompatible with high D/H ratios inferred for the icy satellites (see \textbf{Section 4.1} above), for the water produced as a byproduct thereof would (presumably) equilibrate rapidly with surrounding, low-D, $H_2$ gas. Nonetheless, satellite D/H ratios (with the exception of Enceladus; \textcolor{blue}{Waite et al., 2009}) are all measured from reflectance spectroscopy of surface water-ice, and the assumption that such ice bears the same D/H as that in the interior (comprising the bulk water content of the satellite) remains a dubious one.\\
\indent We consider five astrophysical settings for water production by CO-CH$_4$ conversion:

1. \textit{The circumplanetary disk}: Perhaps CO-CH$_4$ conversion occurred \textit{after} gas was introduced from the circumsolar disk. 

2. \textit{The circumsolar disk}: Perhaps substantial CO-CH$_4$ conversion occurred \textit{before} gas was delivered into circumplanetary disks around the giant planets. 

3. \textit{Meridional flow feeding the circumplanetary disk}: Perhaps efficient conversion can be achieved \textit{during} gas delivery from the circumsolar to circumplanetary disk,. 

4. \textit{Shock processing by incoming planetesimals}: Perhaps the primary building blocks of the satellites are ice-poor planetesimals that hit the disk, converting some CO to water and CH$_4$ and becoming assimilated therein.  

5. \textit{Proto-Jupiter itself}: Perhaps CO-CH$_4$ conversion is accomplished in proto-Jupiter's outermost envelope, and the water produced was subsequently ejected (\textit{e.g.,} via a giant impact), supplying the disk with water. 

We consider these settings consecutively. 

\begin{figure} 
\scalebox{0.85}{\includegraphics{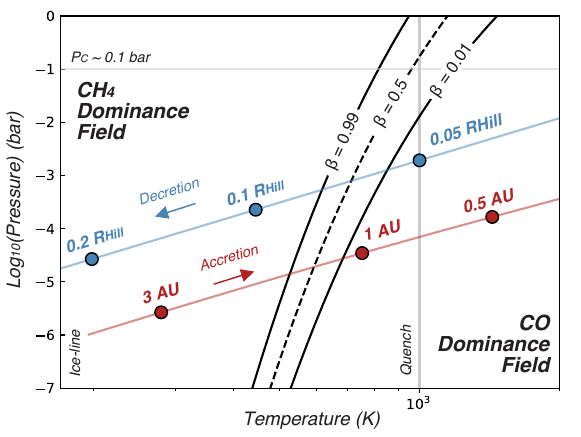}}
\caption{\textbf{Pressure-Temperature profiles of our circumplanetary disk model (blue) and a typical steady-state model for a viscous accretion (circumsolar) disk (red), against the boundary between the CO and CH$_4$ dominance fields.} Contours of $\beta$ represent the fraction of C held in CH$_4$ \textit{at thermodynamic equilibrium}. At and above the quench temperature $T_Q\sim1000K$, pressures are clearly too low for efficient CO-CH$_4$ conversion, and hence water production. See text in \textbf{Section 4.2.1}.}
\label{fig:Figure 9}
\end{figure}

\subsubsection{Disk Conversion: Settings 1 \& 2}
\indent The earliest work on the problem of CO-CH$_4$ conversion in the early SS predates the current dynamical view of the circumplanetary disk, and can be traced to \textcolor{blue}{Prinn \& Fegley (1981)}. It was posited that circumplanetary nebulae were sufficiently hot and dense to permit CH$_4$ and water production at short timescales relative to radial mixing and nebular cooling. This inference was subsequently refuted on grounds that the adiabatic $P-T$ profiles on which it was based are only applicable to pressure-supported atmospheres, not disks that are radially supported by angular momentum. Through a turbulent disk model more akin to those in use today, it was argued that gas pressures in both the  Saturnian \textcolor{blue}{(Mousis et al, 2002)} and Jovian \textcolor{blue}{(Mousis \& Gautier, 2004)} circumplanetary disks were orders of magnitude less that what \textcolor{blue}{Prinn \& Fegley (1981)} had found. Accordingly, CO-CH$_4$ conversion was deemed unfeasible. \\
\indent Three criteria must be met for conversion of CO in circumplanetary disks to be a viable source of water for giant planet satellites: (i) Does the temperature exceed the quench value of $T_Q \sim1000K$ (at the inner edge, where it is hottest)?; (ii) Is the pressure at the quench temperature sufficiently high (on the order of $\sim 0.1$ bar) to lie within the $CH_4$ dominance field?; (iii) If water is produced in the inner disk regions, can it be transported to the birthplaces of the satellites? Given the current paradigm for decreting circumplanetary disks, (iii) appears to be satisfied. As for (i), \textbf{Fig. 1} indicates that disk temperatures exceeding $1000K$ can be achieved within a few $\%$ of the giant planet Hill radius. The presence of such an inner high temperature region is largely insensitive to plausible ranges for disk parameters (note that we have assumed both a relatively low $\alpha\sim 10^{-4}$ and opacity, as controlled by the micron-size metallicity $\mathbb{Z}_\mu$. Hence, all that is left is to consider (ii). \\
\indent In \textbf{Fig. 9}, we plot the $P-T$ profile for our circumplanetary disk model (blue) against the boundary between the CO and CH$_4$ dominance fields, given by
\begin{equation}
P(T) = \sqrt{\frac{\beta}{1-\beta}\frac{(0.77X_{O}-X_{CO})e^{\Delta G^{\circ}(T,P_{0})/RT}}{X_{H2}^{3}}},
\end{equation}
where $X_A$ represents the mole fraction of species A, $\beta$ parametrizes the partitioning of carbon atoms between CH$_4$ and CO \textit{at thermodynamic equilibrium} (\textit{i.e.,} $\beta = X_{CH_4}/X_C$; $X_{CO} = (1-\beta)X_C$), and $\Delta G^{\circ} = 3 G^{\circ}_{H2} +  G^{\circ}_{CO} -  G^{\circ}_{CH4} -  G^{\circ}_{H2O}$ is the change in the Gibbs free energy at a reference pressure $P_0$ of 1 bar for the reaction (\textbf{Eq. 1}), with $G^{\circ}_A = H^{\circ}_A - TS^{\circ}_A$, $H$ and $S$ being the enthalpy and entropy at $P_0$. Moving on, $R = 8.314 $ J$\cdot$mol$^{-1}\cdot$K$^{-1}$ is the ideal gas constant, and we adopt the solar values of $X_{H2} \simeq 0.835$, $X_C \simeq 5.9\times 10^{-4}$, and $X_O\simeq 1.2\times 10^{-3}$ \textcolor{blue}{(Anders \& Grevasse, 1989)}. \\
\indent As mentioned before, the pressure needed for efficient CO-CH$_4$ conversion at the quench temperature of  $T_Q\sim1000K$ is in the vicinity of $P_C \sim0.1$ bar (\textbf{Eq. 2}). While temperatures  $\gtrsim T_Q$ are achieved within $\sim 0.05 R_{Hill}$, the corresponding pressures remain about two orders of magnitude too low for substantial conversion. This conclusion is insensitive to the choice of disk parameters\textemdash higher $T$ leads to higher $P$, but also much higher pressure for entering the CH$_4$ dominance field.
\indent In the circumsolar disk, the same problem is encountered. While criteria (i) and (iii; the circumsolar disk feeds the circumplanetary disk) are met, criterion (ii) is not. In \textbf{Fig. 9}, we also plot the $P-T$ profile for a steady-state viscous accretion disk with a typical stellar accretion rate $\dot{M}_{acc}\simeq 10^{-8} M_{\odot}$/yr $\simeq 3\pi \nu'\Sigma_g'$. Here and onward, primes denote circumsolar disk parameters. The temperature is given by \textbf{Eq. 9}, where here we assume $\mathbb{Z}_\mu' = 0.01$ (\textit{i.e.,} the ISM value). The pressure, as usual, is $P' = \rho_g'c_s'^2$, where $\rho_g' = \Sigma_g'\Omega_k'/\sqrt{2\pi} c_s'$. The circumsolar disk $\alpha'$ is assumed to be $10^{-4}$ as in our circumplanetary disk model. Evidently, while $T\gtrsim T_Q$ is obtained within $\sim0.8$ AU, corresponding pressures are no where close to adequate for CO conversion. \\

\subsubsection{Conversion during Delivery into Circumplanetary Disks: Setting 3}
\indent As mentioned in \textbf{Section 1.2}, it is widely accepted that material delivery into circumplanetary disks proceeds via meridional flows \textcolor{blue}{(Tanigawa et al., 2012; Morbidelli et al., 2014; Teague et al., 2019; Adams \& Batygin, 2025)}. That is, incoming gas and dust falls sub-vertically onto the forming giant planet and its disk from above and below the midplane of the circumstellar disk. While substantial CO-CH$_4$ is unlikely within both the circumsolar and circumplanetary disk (see \textbf{Section 4.2.1} above), we evaluate here its efficacy during the transition between the two. \\
\indent A useful order of magnitude assessment can be made as follows. Suppose we build a giant planet (here we take Jupiter) by inflow through a vertical cylinder of radius $R$, wherein the gas has density $\rho_g$ and is initially close to free-fall at velocity $v_{ff} = \sqrt{2GM_J/R}$ because hydrostatic support is absent. For an accretion timescale $\tau_{acc}$, we have 
\begin{equation}
4\rho_g v_{ff}\tau_{acc}\pi R^2 \sim M_J,
\end{equation}
where the factor of $4$ accounts for inflow from above and below the circumsolar disk, on both sides of the gap carved by Jupiter. Re-arranging, we find
\begin{equation}
\rho_g \sim \frac{\rho_J R_J}{3v_{esc}\tau_{acc}} \left(\frac{R_J}{R}\right)^{3/2},
\end{equation}
where $\rho_J \sim 1330$ kg/m$^3$ is Jupiter's current mean density, $R_J\sim 7\times 10^7$m its present-day radius, and $v_{esc} = v_{ff}\sqrt{R/R_J}\sim 59.5$km/s its escape velocity. Except for highly collimated flows and very short timescales, this yields $\rho_g$ less than or comparable to that in the circumsolar disk (\textit{i.e.,} at $\sim 5$ AU in the viscous accretion disk model in \textbf{Section 4.2.1}, $\rho_g\sim 10^{-7}$ kg/m$^3$). For instance, $\rho_g\sim 5\times  10^{-9}$ kg/m$^3$ for $R\sim 5R_J$ and $\tau_{acc}\sim 0.3$ Myr. With this value of $\rho_g$, the dynamic pressure $\rho_g v_{ff}^2 < 10^{-4}$ bar,  close to four orders of magnitude less than $P_C\sim 0.1$ bar (\textbf{Eq. 2)}. As before for gas within disks, pressure remains the problem. \\
\indent Let us consider a slightly different approach. Assume the model cylinder for the meridional flow has height equivalent to the hydrostatic scale height of the circumsolar disk $H_g$ (after all, this is where incoming gas and dust is thought to originate; see \textbf{Section 1.2}). The cylinder volume, then, is $V_c \sim H_g R^2$. While gas travels down the cylinder at $\sim v_{ff}$ for most of its journey, it begins with a velocity approximated by the radial, viscous velocity of gas of the circumsolar disk, given by \textbf{Eq. 10}. This yields, upon substitution of \textbf{Eq. 8}: 
\begin{equation}
v_{r,g}' = \alpha'\left(\frac{H_g}{r'}\right)^2 r' \Omega_k' \left(1 + \frac{dln\Sigma_g'}{dlnr'}\right),
\end{equation}
where we assume  $\alpha'\sim 10^{-4}$, $r'\sim 5$ AU, and the aspect ratio $H_g/r' \sim 0.05$ (note that primes on disk parameters distinguish them from those in the circumplanetary disk).  The maximum slope in $\Sigma_g'$ across the carved gap is given by \textcolor{blue}{(Crida et al., 2006)} in the limit of vanishing viscosity
\begin{equation}
\frac{dlog_{10}(\Sigma_g')}{dr'} = \frac{1.3}{ H_g}, 
\end{equation}
which yields 
\begin{equation}
v_{r,g}' = \alpha'\left(\frac{H_g}{r'}\right)^2 r' \Omega_k' \left[1 + \frac{3r'}{H_g}\right].
\end{equation}
With this, we can compute the mass flow rate into the cylinder $\dot{M}_C$, expressed as
\begin{equation}
\dot{M}_C = 2\pi r' v_{r,g}' \Sigma_{g,a}',
\end{equation}
where $\Sigma_{g,a}'\sim 100$ kg/m$^2$ is the surface density of the MRI-active disk layer driving the viscous flow \textcolor{blue}{(Morbidelli et al., 2014)}. \\
\indent Assuming a steady-state flow through the cylinder, the mass exiting the it close to the circumplanetary disk midplane must also be $\dot{M}_C$. Since $v_{ff}>v_{r,g}'$, we have relatively slow-moving and dense gas at the top of the cylinder, and fast-moving, thin gas throughout most of its volume. Defining the free-fall timescale $\tau_{ff}\sim H_g/v_{ff}$, the total mass of gas in the cylinder at any given time can be approximated as
\begin{equation}
M_C \sim \dot{M}_C\tau_{ff} \sim \frac{2\sqrt{5}\pi r' v_{r,g}'\Sigma_{g,a}'H_g}{v_{esc}},
\end{equation}
where we have assumed, as before, that $R\sim 5R_J$. The characteristic gas density in the cylinder, then, is $\rho_g\sim M_C/V_C$, from which we get the dynamic pressure
\begin{equation}
\rho_gv_{ff}^2 \sim \frac{2\sqrt{5}\pi r'v_{r,g}'\Sigma_{g,a}'v_{esc}}{125R_J^2}, 
\end{equation}
which evaluates to $\sim 2 \times 10^{-4}$ bar, once again far less than what is needed for CO-CH$_4$ conversion to take place at a meaningful rate. \\
\indent Given that $v_{ff} \gg c_s$ even at the disk inner edge, a shockfront could form at the centrifugal radius of infall \textcolor{blue}{(Szulágyi, 2017)}, generating (post-shock) pressures and temperatures favorable for efficient CO-CH$_4$ conversion. Nonetheless, the dynamics of infall, specifically angular momentum transfer, remain poorly understood. Accordingly $v_{ff}$ may grossly overestimate the gas velocity at ``impact" with the disk (note that our use of $v_{ff}$ in the calculations above is conservative). The problem is complicated by the unconstrained effect of opacity, or more generally, the amount of cooling that occurs. A detailed investigation into the $P-T$ conditions characterizing meridional flows, beyond the scope of this work, is an enticing avenue for future research. 

\subsubsection{Shock Conversion by Planetesimals: Setting 4}
\indent Planetesimals on heliocentric orbits can be assimilated into circumplanetary disks via gas drag \textcolor{blue}{(Fujita et al., 2013; Tanigawa et al., 2014; D'Angelo \& Podolak; 2015; Suetsugu et al., 2016; Suetsugu \& Ohtsuki, 2017).} In doing so, such planetesimals typically encounter a gas mass on the same order as their own masses. At $r\sim0.4 R_{Hill}$, the outer edge of the circumplanetary disk as carved by tidal truncation \textcolor{blue}{(Martin \& Lubow, 2011)}, $c_s\sim 550$ m/s in our disk model, far smaller than the orbital velocity $v_k\sim2500$ km/s. As such, most planetesimal-disk encounters will be supersonic (especially for planetesimals approaching the giant planet on retrograde trajectories\textemdash that is, against the direction of disk rotation). At their shock fronts, sufficiently high pressures and temperatures can induce CO-CH$_4$ conversion.\\
\indent While a promising setting for water production, the total mass of water created is of the same order as the fraction of disk mass that is initially CO, multiplied by the total incoming mass of planetesimals. This is likely far lower than what is needed. Indeed, the planetesimal flux is expected to be low during the nebular epoch. Assuming conservatively that ten Earth masses of planetesimals were assimilated, and a CO gas mass fraction of $\sim0.1\%$, the mass of water produced only amounts to only $\sim10\%$ the mass of Ganymede, Callisto, or Titan ($\sim 10^{23}$ kg). Moreover, this mechanism appeals to coincidence\textemdash there is no clear reason to expect from this process, even if more efficient than presently thought, that the ice-to-rock ratios in the three satellites would be close to unity. 

\subsubsection{Supply from Jupiter: Setting 5}
\indent Giant impacts may have occurred during the formation of Jupiter and Saturn, leading to some ejection of their atmospheres. There is currently a disagreement between the gravity field of Jupiter inferred from Juno data and the observed excess of heavy elements in its atmosphere, and a giant impact is suggested as a possible explanation thereof \textcolor{blue}{(Helled et al., 2022)}. 
Owing to CO-CH$_4$ conversion in its outer envelope \textcolor{blue}{(Cavalie et al., 2023)}, the ice-to-rock ratio in Jupiter's atmosphere might be the same as in Ganymede and Callisto, but this is not known since any rock component forms clouds deep below the observable region. It is unlikely that ejection from Jupiter (or Saturn) is a dynamically realistic explanation of the water-rich satellites. Nonetheless, it is worth noting that the total ice and rock in the Galilean satellites is only of order $\sim1\%$ or less of the total mass of heavy elements in Jupiter, and this amount is present in the outermost $\sim 10$ to $20\%$ of its present radius \textcolor{blue}{(Helled \& Stevenson, 2024)}.\\

\section{Concluding Remarks}
\indent In this work, we show that ice accumulation at the decretion disk ice-line is a natural consequence of outward water vapor advection within it, coupled with inward icy pebble drift beyond it. For typical disk parameters, we find that such accumulation results in a solid reservoir enriched in water-ice by a factor of $\sim 2$  to $3$ compared to solids across the disk as a whole (\textit{i.e.,} the ``background"). An increase in this factor accompanies that in the turbulent diffusivity, reflecting either a decrease in the turbulent Schmidt number or an increase in the Shakura-Sunyaev $\alpha$ parameter. We propose this ``cold finger" mechanism explains, at least in part, the anomalously water-rich compositions of the SS giant planet satellites Ganymede, Callisto, and Titan. Furthermore, it is compatible with their inferred high D/H ratios. While we have been unable to identify a compelling setting for substantial water production via CO-CH$_4$ conversion, we recognize the remaining uncertainties regarding the dynamics (and thermodynamics) of giant planet runaway gas accretion and circumplanetary disk evolution, marking fertile ground for research. Investigating the extent of water accumulation at the ice-line within a global and evolving disk model also constitutes a worthwhile avenue for future work. 

\section*{Appendix}
\subsection*{Length-Scale for\\ Pebble Cloud Collapse}
\indent Satellitesimals in circumplanetary disks, like planetesimals in circumstellar disks, are thought to form from the gravitational self-collapse of pebble clouds. Importantly, collapse can only be realized if the said clouds (of length scale $\lambda$) are characterized by a sufficiently high metallicity, as may be achieved through vortices, zonal flows, or dust trapping due to gas-dust instabilities (\textit{e.g.,} the streaming instability; \textcolor{blue}{Youdin \& Goodman, 2005; Squire \& Hopkins, 2018}). This must be so for them to be stable against giant planet tidal disruption, and contract under self-gravity at timescales considerably less than the those associated with (i) ``pressure," or kinetic, support (\textit{i.e.,} $\sim \lambda/\sigma_{pb}$; where $\sigma_{pb}$ is the pebble velocity dispersion), (ii) Keplerian shear (\textit{i.e.,} $\sim \Omega_k^{-1}$), and (iii) turbulent diffusion (\textit{i.e.,} $\sim \lambda^2/D$) \textcolor{blue}{(\textit{e.g.,} Armitage, 2020; Gerbig et al., 2020; Klahr \& Schreiber, 2020).}\\
\indent In our simulations, the pebble metallicity of the ice-rich reservoir at steady-state only exceeds the initial, ISM value of $\sim1\%$ by a factor of a few (see \textbf{Figs. 6, 7 \& 8}). While this appears an order of magnitude too small for satellitesimal formation to occur, bear in mind that this applies to an annulus of width $\Delta r =0.01 R_{Hill}$ that is much larger than the length scale $\lambda$ associated with the pebble clouds envisioned. The purpose of our work is merely to show that enhanced ice-to-rock ratios can be obtained by dust in the vicinity of the ice-line, and we remain agnostic to the local processes occurring \textit{within} $\Delta r$ that can lead to high metallicities within a patch of disk area $\lambda^2$. \\
\indent Estimations of $\lambda$ follow from the criterion for pebble cloud collapse. Here, we briefly review two complementary approaches towards that criterion. In both, consideration of timescales renders the derivation of the criterion amenable to analytic treatment. The first calls upon the Toomre $Q$, as applied to a pebble layer \textcolor{blue}{(Toomre, 1964; Goldreich \& Ward, 1973; Youdin \& Shu, 2002; Batygin \& Morbidelli, 2020)}, obtained by equating the cloud free-fall timescale to the timescales for pressure support and shear (this assumes the pebble layer is a fluid, which sound speed is equivalent to the pebble velocity dispersion). The second is a Jeans-length approach taken by \textcolor{blue}{Klahr \& Schreiber (2020)}, analogous to that for star formation (\textit{i.e.,} molecular cloud collapse; \textcolor{blue}{Jeans, 1902}), obtained by equating the free-fall timescale (or contraction timescale, accounting for pebble drag during collapse), to the timescale of turbulent diffusion. 

\subsection*{Toomre-Q Approach}
\indent The Toomre $Q$  for a pebble layer takes the form
\begin{equation}
Q_{pb} = \frac{h_{pb} \Omega_k^2}{\pi G \Sigma_{pb}},
\end{equation}
where $h_{pb}\sim \sigma_{pb}/\Omega_K$ represents the scale height of the pebble sub-disk, and $\Sigma_{pb}$ the pebble surface density. Following \textcolor{blue}{Batygin \& Morbidelli (2020)}, we derive $h_{pb}$ assuming it is limited by the kinetic energy held in turbulent eddies. Note that the expression for the scale height of the solid sub-disk in \textbf{Eq. 5} is valid only in the limit where the dust mass is negligible (\textit{i.e.,} $\mathbb{Z}_{pb} <<0.1$). The vertically integrated turbulence kinetic energy density is given by $\mathcal{E}_t = \alpha c_s^2\Sigma_g/2$. Setting $f_t\mathcal{E}_t$ equal to the gravitational potential energy density of the pebble layer $\mathcal{E}_G = \Sigma_{pb}\Omega_k^2 h_{pb}^2$, where $f_t\sim 0.1$ accounts for the fact that only a fraction of the kinetic energy is expended on vertical stirring of pebbles and the expected suppression of turbulence by high metallicities \textcolor{blue}{(Dullemond \& Penzlin, 2018; Lin, 2019)}, yields
\begin{equation}
h_{pb} = \frac{c_s}{\Omega_k}\sqrt{\frac{f_t\alpha}{2\mathbb{Z}_{pb}}}.
\end{equation}
Here, the pebble metallicity $\mathbb{Z}_{pb} = \Sigma_{pb}/\Sigma_g$ as in our dust evolution model. Collapse ensues for $Q_{pb}\lesssim1$, or
\begin{equation}
\mathbb{Z}_{pb}\gtrsim \left(\frac{c_s\Omega_k}{\pi G \Sigma_g}\sqrt{\frac{f_t\alpha}{2}}\right)^{2/3} = Q^{2/3}\left(\frac{f_t\alpha}{2}\right)^{1/3},
\end{equation}
where in the second equality, we substituted for the classic Toomre $Q = c_s\Omega_k/\pi G \Sigma_g$ for the stability of disk gas. Evaluating the above expression beyond the ice-line in our model (\textit{i.e.,} bin 2) yields $\mathbb{Z}_{pb}\gtrsim 0.2$ for satellitesimal collapse to occur. High metallicity ($\mathbb{Z}_{pb}>> 0.01$), if obtained, ensures that pebble layers/clouds remain intact in the face of Kelvin-Helmholtz instabilities that (i) arise at the interface between the layer surface orbiting at Keplerian velocity and the sub-Keplerian gas \textcolor{blue}{(Cuzzi et al., 1993)}, and would otherwise prevent the pebble sedimentation sufficient for collapse \textcolor{blue}{(Sekiya, 1998)}. \\
\indent To estimate $\lambda$, consider the dispersion relation for a self-gravitating particle disk in a Keplerian potential \textcolor{blue}{(Armitage, 2020; Batygin \& Morbidelli, 2020)}:
\begin{equation}
\omega(k)^2 = \Omega_k^2 +(\sigma_{pb} k)^2 - 2\pi G \Sigma_{pb}|k|,
\end{equation}
where $\omega$ and $k$ are the temporal frequency and spatial wavenumber of the perturbations to the disks' equilibrium. Setting $d\omega^2/dk = 0$, we find that the most rapidly growing mode is characterized by 
\begin{equation}
k_{crit} = \frac{\pi G\Sigma_{pb}}{(h_{pb}\Omega_k)^2} = \frac{2\pi G \Sigma_g \mathbb{Z}_{pb}^2}{f_t \alpha c_s^2}.
\end{equation}
Accordingly, the length scale associated with cloud collapse $\lambda$ is given by
\begin{equation}
\lambda \sim \frac{2\pi}{k_{crit}}, 
\end{equation}
which evaluates to $\sim0.002 R_{Hill}$, smaller ($\sim5$ times) than $\Delta r$ as expected. Importantly, $\lambda$ is likely smaller than the scale of the pebble layer itself\textemdash the layer, having reached $\mathbb{Z}_{pb}\sim 0.2$, promotes the formation of satellitesimals within it.

\subsection*{Jeans-Length Approach}
\indent In the Jeans criterion for collapse, the high metallicity prerequisite takes the form of a threshold pebble cloud density, above which the cloud withstands disruption from the giant planet. This density is obtained by asserting that the cloud's gravitational influence on its constituent pebbles exceeds that of the planet\textemdash that is, the pebbles lies within the Hill sphere of the cloud (center of mass). This yields
\begin{equation}
\rho_{Hill} = \frac{9M_J}{4\pi r^3},
\end{equation}
where $M_J$ is the mass of Jupiter (our representative giant planet in this work; see \textbf{Section 2}).  To link $\rho_{Hill}$ to $\mathbb{Z}_{pb}$ from the Toomre $Q$ approach above, note that
\begin{equation}
\rho_{Hill}\sim \frac{\Sigma_{pb}}{\sqrt{2\pi}h_{pb}} = \frac{\Sigma_{pb}}{\sqrt{2\pi}h} \sqrt{\frac{2\mathbb{Z}_{pb}}{f_t \alpha}} = \rho_g \mathbb{Z}_{pb} \sqrt{\frac{2\mathbb{Z}_{pb}}{f_t \alpha}},
\end{equation}
where we have made use of \textbf{Eq. 39}, and the relations $h = c_s/\Omega_k$ and $\Sigma_{pb} = \mathbb{Z}_{pb} \Sigma_g$. Re-arranging for $\mathbb{Z}_{pb}$, we obtain
\begin{equation}
\mathbb{Z}_{pb} \sim \left(\frac{\rho_{Hill}}{\rho_g}\sqrt{\frac{f_t\alpha}{2}}\right)^{2/3}.
\end{equation}
Upon evaluation at the ice-line (\textit{i.e.,} in bin 2), we obtain $\mathbb{Z}_{pb} \sim 0.36$, consistent with the condition $\mathbb{Z}_{pb}\gtrsim 0.2$ above.\\
\indent Once the cloud achieves a density $\gtrsim \rho_{Hill}$, the viability of collapse depends on the competition between contraction (\textit{i.e.,} gravity) and turbulent diffusion. The timescale for the former is given by
\begin{equation}
\tau_c = \tau_{ff}\left(1+\frac{8\Omega_k\tau_{ff}}{3\pi^2 St}\right),
\end{equation}
where $St$ is the characteristic pebble Stokes number (see \textbf{Eq. 15}), and the free-fall timescale $\tau_{ff}$ is
\begin{equation}
\tau_{ff} = \sqrt{\frac{3\pi}{32 G\rho_{Hill}}}.
\end{equation}
The second term in $\tau_c$ represents the contribution of gas drag on the collapsing pebbles (leading to their infall at terminal velocity), and dominates for $St<<1$, as in our model (see \textbf{Section 2.3.1}). The timescale for turbulent diffusion $\tau_D$, assumed to be inherited from the Kolmogorov cascade of global turbulence, takes the form
\begin{equation}
\tau_D = \frac{\lambda^2}{D} = \frac{\lambda^2 }{\delta c_s h},
\end{equation}
where we have expressed the turbulent diffusivity in terms of the dimensionless diffusivity $\delta$ (analogous to $\alpha$ for $\nu = Sc D$). Assuming $Sc\sim 1$, the diffusivity inherited at the scale of satellitesimal formation from the turbulent cascade $\delta(\lambda)$ is given by \textcolor{blue}{(Schreiber \& Klahr, 2018)}
\begin{equation}
\delta(\lambda)\sim \frac{\alpha^{1/3}\rho_g}{\rho_g + \rho_{Hill}} \left(\frac{\lambda}{h}\right)^{4/3}.
\end{equation}
Substituting the above into the expression for $\tau_D$ and setting $\tau_C = \tau_D$ yields
\begin{equation}
\lambda \sim \frac{\sqrt{\alpha}h}{27}\left[\frac{\rho_g}{St (\rho_g + \rho_{Hill})}\right]^{3/2}.
\end{equation}
This evaluates to $\sim 7\times 10^{-4}  R_{Hill}$ for $St\sim 10^{-3}$ (corresponding to pebbles in the few tens of mm at the ice-line). As deduced from the Toomre $Q$ approach, $\lambda$ is far smaller (here close to $\sim 15$ times) than the bin width $\Delta r = 0.01R_{Hill}$. Assuming $\lambda\sim 0.001R_{Hill}$, the mass of a satellitesimal formed would be $\sim \mathbb{Z}_{pb}\Sigma_g \lambda^2$. For $\mathbb{Z}_{pb}\sim 0.3$, this evaluates to $\sim 2\times 10^{-19}$ kg. For a satellitesimal bulk density of $\sim 2000$ kg/m$^3$, this implies a radius of $\sim 130$km, consistent with inferences of planetesimal initial sizes from both theory \textcolor{blue}{(Klahr \& Schreiber, 2020; Gerbig \& Li. 2023)} and observation \textcolor{blue}{(Delbo et al., 2017)}. \\

\subsection*{Brief note on\\ Pebble Cloud Formation}
\indent Several mechanisms have been proposed to engender concentrated pebble clouds, primarily in the context of planetesimal formation in circumstellar accretion disks. These include a plethora of two-fluid instabilities between gas and dust, most notably the streaming instability \textcolor{blue}{(Youdin \& Goodman, 2005; Takahashi \& Inutsuka, 2014; Squire \& Hopkins, 2018)}, as well as the formation of pressure bumps \textcolor{blue}{(\textit{e.g.,} Taki et al., 2016; Jiang \& Ormel, 2023)}, which may themselves promote the said instabilities. Notably, the ice-line has been regarded as a favorable site for pressure bump formation \textcolor{blue}{(\textit{e.g.,} Kretke \& Lin, 2007; Brauer et al., 2008a; Bitsch et al., 2015; Charnoz et al., 2021; Müller et al., 2021)}. For decretion disks specifically, a qualitatively different picture for pebble accumulation has been presented by \textcolor{blue}{Batygin \& Morbidelli (2020)}. As the direction of radial gas advection is opposite that of inward pebble drift due to headwind drag, the disk metallicity is expected to build up globally over time in pebbles for which $v_{r,d}\sim0$ (\textit{i.e.,} in pebbles whose size/$St$ is such that the two terms in \textbf{Eq. 16} cancel out).  \\

\section*{Acknowledgments}
\indent We extend our gratitude to Konstantin Batygin for insightful discussions on dust coagulation at the early stages of this work, as well as suggestions for improving the clarity of our manuscript. We also thank two anonymous reviewers for their constructive feedback, which strengthened our proposed scenario for icy satellite formation, and  editor Brian Jackson for careful and prompt editorial handling. This work was supported by the Caltech Center for Comparative Planetary Evolution (3CPE). 

\section*{References}

Adams, F. C., \& Batygin, K. (2025). General Analytic Solutions for Circumplanetary Disks during the Late Stages of Giant Planet Formation. \textit{Publications of the Astronomical Society of the Pacific}, \textit{137}(5), 054401.\\

Altwegg, K., Balsiger, H., Bar-Nun, A., Berthelier, J. J., Bieler, A., Bochsler, P., ... \& Wurz, P. (2015). 67P/Churyumov-Gerasimenko, a Jupiter family comet with a high D/H ratio. \textit{Science}, \textit{347}(6220), 1261952.\\

Anders, E., \& Grevesse, N. (1989). Abundances of the elements: Meteoritic and solar. \textit{Geochimica et Cosmochimica acta}, \textit{53}(1), 197-214.\\

Armitage, P. J. (2020). \textit{Astrophysics of planet formation}. Cambridge University Press.\\

Asplund, M., Grevesse, N., Sauval, A. J., \& Scott, P. (2009). The chemical composition of the Sun. \textit{Annual review of astronomy and astrophysics}, \textit{47}(2009), 481-522.\\

Balbus, S. A., \& Hawley, J. F. (1991). A powerful local shear instability in weakly magnetized disks. I-Linear analysis. II-Nonlinear evolution. \textit{Astrophysical Journal, Part 1 (ISSN 0004-637X), vol. 376, July 20, 1991, p. 214-233.}, \textit{376}, 214-233.\\

Bate, M. R., Lubow, S. H., Ogilvie, G. I., \& Miller, K. A. (2003). Three-dimensional calculations of high-and low-mass planets embedded in protoplanetary discs. \textit{Monthly Notices of the Royal Astronomical Society}, \textit{341}(1), 213-229.\\

Batygin, K., \& Adams, F. C. (2025). Determination of Jupiter’s primordial physical state. \textit{Nature Astronomy}, 1-10.\\

Batygin, K., \& Morbidelli, A. (2020). Formation of giant planet satellites. \textit{The Astrophysical Journal}, \textit{894}(2), 143.\\

Batygin, K., \& Morbidelli, A. (2022). Self-consistent model for dust-gas coupling in protoplanetary disks. \textit{Astronomy \& Astrophysics}, \textit{666}, A19.\\

Batygin, K., \& Morbidelli, A. (2023). Formation of rocky super-earths from a narrow ring of planetesimals. \textit{Nature Astronomy}, \textit{7}(3), 330-338.\\

Bézard, B., Nixon, C. A., Vinatier, S., Lellouch, E., Greathouse, T., Giles, R., ... \& Jolly, A. (2024). The D/H ratio in Titan’s acetylene from high spectral resolution IRTF/TEXES observations. \textit{Icarus}, \textit{421}, 116221.\\

Birnstiel, T., Ormel, C. W., \& Dullemond, C. P. (2011). Dust size distributions in coagulation/fragmentation equilibrium: numerical solutions and analytical fits. \textit{Astronomy \& Astrophysics}, \textit{525}, A11.\\

Birnstiel, T., Klahr, H., \& Ercolano, B. (2012). A simple model for the evolution of the dust population in protoplanetary disks. \textit{Astronomy \& Astrophysics}, \textit{539}, A148.\\

Bitsch, B., Morbidelli, A., Lega, E., \& Crida, A. (2014). Stellar irradiated discs and implications on migration of embedded planets-II. Accreting-discs. \textit{Astronomy \& Astrophysics}, \textit{564}, A135.\\

Bitsch, B., Johansen, A., Lambrechts, M., \& Morbidelli, A. (2015). The structure of protoplanetary discs around evolving young stars. \textit{Astronomy \& Astrophysics}, \textit{575}, A28.\\

Blum, J., \& Münch, M. (1993). Experimental investigations on aggregate-aggregate collisions in the early solar nebula. \textit{Icarus}, \textit{106}(1), 151-167.\\

Bockelée-Morvan, D., Biver, N., Swinyard, B., de Val-Borro, M., Crovisier, J., Hartogh, P., ... \& Walker, H. (2012). Herschel measurements of the D/H and 16O/18O ratios in water in the Oort-cloud comet C/2009 P1 (Garradd). \textit{Astronomy \& Astrophysics}, \textit{544}, L15.\\

Bockelée-Morvan, D., Calmonte, U., Charnley, S., Duprat, J., Engrand, C., Gicquel, A., ... \& Wirström, E. (2015). Cometary isotopic measurements. \textit{Space science reviews}, \textit{197}(1), 47-83.\\

Bockelée-Morvan, D., Gautier, D., Lis, D. C., Young, K., Keene, J., Phillips, T., ... \& Wootten, A. (1998). Deuterated water in comet C/1996 B2 (Hyakutake) and its implications for the origin of comets. \textit{Icarus}, \textit{133}(1), 147-162.\\

Bohlin, R. C., Savage, B. D., \& Drake, J. F. (1978). A survey of interstellar HI from L-alpha absorption measurements. II. \textit{Astrophysical Journal, Part 1, vol. 224, Aug. 15, 1978, p. 132-142. Research supported by the Lockheed Independent Research Program;}, \textit{224}, 132-142.\\

Brauer, F., Henning, T., \& Dullemond, C. P. (2008a). Planetesimal formation near the snow line in MRI-driven turbulent protoplanetary disks. \textit{Astronomy \& Astrophysics}, \textit{487}(1), L1-L4.\\

Brauer, F., Dullemond, C. P., \& Henning, T. (2008b). Coagulation, fragmentation and radial motion of solid particles in protoplanetary disks. \textit{Astronomy \& Astrophysics}, \textit{480}(3), 859-877.\\

Brown, M. E., Trumbo, S. K., Davis, M. R., \& Chandra, S. (2025). Deuterated water ice on the satellites of Saturn. \textit{The Planetary Science Journal}, \textit{6}(10), 229.\\

Brown, P. D., \& Millar, T. J. (1989). Models of the gas–grain interaction–deuterium chemistry. \textit{Monthly Notices of the Royal Astronomical Society}, \textit{237}(3), 661-671.\\

Caffau, E., Ludwig, H. G., Steffen, M., Freytag, B., \& Bonifacio, P. (2011). Solar chemical abundances determined with a CO5BOLD 3D model atmosphere. \textit{Solar Physics}, \textit{268}(2), 255-269.\\

Canup, R. M., \& Ward, W. R. (2002). Formation of the Galilean Satellites: Conditions ofAccretion. \textit{The Astronomical Journal}, \textit{124}(6), 3404.\\

Cavalié, T., Lunine, J., \& Mousis, O. (2023). A subsolar oxygen abundance or a radiative region deep in Jupiter revealed by thermochemical modelling. \textit{Nature Astronomy}, \textit{7}(6), 678-683.\\

Charnoz, S., Avice, G., Hyodo, R., Pignatale, F. C., \& Chaussidon, M. (2021). Forming pressure traps at the snow line to isolate isotopic reservoirs in the absence of a planet. \textit{Astronomy \& Astrophysics}, \textit{652}, A35.\\

Ciesla, F. J., \& Cuzzi, J. N. (2006). The evolution of the water distribution in a viscous protoplanetary disk. \textit{Icarus}, \textit{181}(1), 178-204.\\

Clark, R. N., Brown, R. H., Cruikshank, D. P., \& Swayze, G. A. (2019). Isotopic ratios of Saturn's rings and satellites: Implications for the origin of water and Phoebe. \textit{Icarus}, \textit{321}, 791-802.\\

Crida, A., Morbidelli, A., \& Masset, F. (2006). On the width and shape of gaps in protoplanetary disks. \textit{Icarus}, \textit{181}(2), 587-604.\\

Cuzzi, J. N., Dobrovolskis, A. R., \& Champney, J. M. (1993). Particle-gas dynamics in the midplane of a protoplanetary nebula. \textit{Icarus}, \textit{106}(1), 102-134.\\

Cuzzi, J. N., \& Zahnle, K. J. (2004). Material enhancement in protoplanetary nebulae by particle drift through evaporation fronts. \textit{The Astrophysical Journal}, \textit{614}(1), 490.\\

D'Angelo, G., Henning, T., \& Kley, W. (2002). Nested-grid calculations of disk-planet interaction. \textit{Astronomy \& Astrophysics}, \textit{385}(2), 647-670.\\

D’Angelo, G., \& Podolak, M. (2015). Capture and evolution of planetesimals in circumjovian disks. \textit{The Astrophysical Journal}, \textit{806}(2), 203.\\

Delbo’, M., Walsh, K., Bolin, B., Avdellidou, C., \& Morbidelli, A. (2017). Identification of a primordial asteroid family constrains the original planetesimal population. \textit{Science}, \textit{357}(6355), 1026-1029.\\

Dipierro, G., Laibe, G., Alexander, R., \& Hutchison, M. (2018). Gas and multispecies dust dynamics in viscous protoplanetary discs: the importance of the dust back-reaction. \textit{Monthly Notices of the Royal Astronomical Society}, \textit{479}(3), 4187-4206.\\

Drażkowska, J., \& Alibert, Y. (2017). Planetesimal formation starts at the snow line. \textit{Astronomy \& Astrophysics}, \textit{608}, A92.\\

Drouart, A., Dubrulle, B., Gautier, D., \& Robert, F. (1999). Structure and transport in the solar nebula from constraints on deuterium enrichment and giant planets formation. \textit{Icarus}, \textit{140}(1), 129-155.\\

Dubrulle, B., Morfill, G., \& Sterzik, M. (1995). The dust subdisk in the protoplanetary nebula. \textit{icarus}, \textit{114}(2), 237-246.\\

Dullemond, C. P., Birnstiel, T., Huang, J., Kurtovic, N. T., Andrews, S. M., Guzmán, V. V., ... \& Ricci, L. (2018). The disk substructures at high angular resolution project (DSHARP). VI. Dust trapping in thin-ringed protoplanetary disks. \textit{The Astrophysical Journal Letters}, \textit{869}(2), L46.\\

Dullemond, C. P., \& Penzlin, A. B. T. (2018). Dust-driven viscous ring-instability in protoplanetary disks. \textit{Astronomy \& Astrophysics}, \textit{609}, A50.\\

Fujita, T., Ohtsuki, K., Tanigawa, T., \& Suetsugu, R. (2013). Capture of planetesimals by gas drag from circumplanetary disks. \textit{The Astronomical Journal}, \textit{146}(6), 140.\\

Geiss, J., \& Gloeckler, G. (1998). Abundances of deuterium and helium-3 in the protosolar cloud. \textit{Space Science Reviews}, \textit{84}(1), 239-250.\\

Geiss, J., \& Reeves, H. (1981). Deuterium in the solar system. \textit{Astronomy and Astrophysics, vol. 93, no. 1-2, Jan. 1981, p. 189-199. Research supported by the Swiss National Science Foundation.}, \textit{93}, 189-199.\\

Gerbig, K., \& Li, R. (2023). Planetesimal initial mass functions following diffusion-regulated gravitational collapse. \textit{The Astrophysical Journal}, \textit{949}(2), 81.\\

Gerbig, K., Murray-Clay, R. A., Klahr, H., \& Baehr, H. (2020). Requirements for gravitational collapse in planetesimal formation—The impact of scales set by Kelvin–Helmholtz and nonlinear streaming instability. \textit{The Astrophysical Journal}, \textit{895}(2), 91.\\

Ghosh, P., \& Lamb, F. K. (1979). Accretion by rotating magnetic neutron stars. II-Radial and vertical structure of the transition zone in disk accretion. \textit{Astrophysical Journal, Part 1, vol. 232, Aug. 15, 1979, p. 259-276.}, \textit{232}, 259-276.\\

Goldreich, P., \& Ward, W. R. (1973). The formation of planetesimals. \textit{Astrophysical Journal, Vol. 183, pp. 1051-1062 (1973)}, \textit{183}, 1051-1062.\\

Grasset, O., Dougherty, M. K., Coustenis, A., Bunce, E. J., Erd, C., Titov, D., ... \& Van Hoolst, T. (2013). JUpiter ICy moons Explorer (JUICE): An ESA mission to orbit Ganymede and to characterise the Jupiter system. \textit{Planetary and Space Science}, \textit{78}, 1-21.\\

Gundlach, B., Kilias, S., Beitz, E., \& Blum, J. (2011). Micrometer-sized ice particles for planetary-science experiments–I. Preparation, critical rolling friction force, and specific surface energy. \textit{Icarus}, \textit{214}(2), 717-723.\\

Gundlach, B., \& Blum, J. (2014). The stickiness of micrometer-sized water-ice particles. \textit{The Astrophysical Journal}, \textit{798}(1), 34.\\

Güttler, C., Blum, J., Zsom, A., Ormel, C. W., \& Dullemond, C. P. (2010). The outcome of protoplanetary dust growth: pebbles, boulders, or planetesimals?-I. Mapping the zoo of laboratory collision experiments. \textit{Astronomy \& Astrophysics}, \textit{513}, A56.\\

Helled, R., \& Stevenson, D. (2017). The fuzziness of giant planets’ cores. \textit{The Astrophysical Journal Letters}, \textit{840}(1), L4.\\

Helled, R., Stevenson, D. J., Lunine, J. I., Bolton, S. J., Nettelmann, N., Atreya, S., ... \& Hubbard, W. B. (2022). Revelations on Jupiter's formation, evolution and interior: Challenges from Juno results. \textit{Icarus}, \textit{378}, 114937.\\

Herbst, E. (1995). Chemistry in the interstellar medium. \textit{Annual Review of Physical Chemistry}, \textit{46}(1), 27-54.\\

Homma, T., Ohtsuki, K., Maeda, N., Suetsugu, R., Machida, M. N., \& Tanigawa, T. (2020). Delivery of pebbles from the protoplanetary disk into circumplanetary disks. \textit{The Astrophysical Journal}, \textit{903}(2), 98.\\

Jeans, J. H. (1902). I. The stability of a spherical nebula. \textit{Philosophical Transactions of the Royal Society of London. Series A, Containing Papers of a Mathematical or Physical Character}, \textit{199}(312-320), 1-53.\\

Jiang, H., \& Ormel, C. W. (2023). Efficient planet formation by pebble accretion in ALMA rings. \textit{Monthly Notices of the Royal Astronomical Society}, \textit{518}(3), 3877-3900.\\

Klahr, H., \& Schreiber, A. (2020). Turbulence sets the length scale for planetesimal formation: Local 2D simulations of streaming instability and planetesimal formation. \textit{The Astrophysical Journal}, \textit{901}(1), 54.\\

Kornet, K., Stepinski, T. F., \& Różyczka, M. (2001). Diversity of planetary systems from evolution of solids in protoplanetary disks. \textit{Astronomy \& Astrophysics}, \textit{378}(1), 180-191.\\

Krapp, L., Kratter, K. M., Youdin, A. N., Benítez-Llambay, P., Masset, F., \& Armitage, P. J. (2024). A thermodynamic criterion for the formation of Circumplanetary Disks. \textit{The Astrophysical Journal}, \textit{973}(2), 153.\\

Kretke, K. A., \& Lin, D. N. C. (2007). Grain retention and formation of planetesimals near the snow line in MRI-driven turbulent protoplanetary disks. \textit{The Astrophysical Journal}, \textit{664}(1), L55.\\

Kuskov, O. L., \& Kronrod, V. A. (2001). Core sizes and internal structure of Earth's and Jupiter's satellites. \textit{Icarus}, \textit{151}(2), 204-227.\\

Lacy, J. H., Carr, J. S., Evans, N. J., Baas, F., Achtermann, J. M., \& Arens, J. F. (1991). Discovery of interstellar methane-Observations of gaseous and solid CH4 absorption toward young stars in molecular clouds. \textit{Astrophysical Journal, Part 1 (ISSN 0004-637X), vol. 376, Aug. 1, 1991, p. 556-560. Research supported by Texas Advanced Research Program.}, \textit{376}, 556-560.\\

Lécluse, C., \& Robert, F. (1994). Hydrogen isotope exchange reaction rates: Origin of water in the inner solar system. \textit{Geochimica et Cosmochimica Acta}, \textit{58}(13), 2927-2939.\\

Lellouch, E., Bézard, B., Fouchet, T., Feuchtgruber, H., Encrenaz, T., \& de Graauw, T. (2001). The deuterium abundance in Jupiter and Saturn from ISO-SWS observations. \textit{Astronomy \& Astrophysics}, \textit{370}(2), 610-622.\\

Lin, M. K. (2019). Dust settling against hydrodynamic turbulence in protoplanetary discs. \textit{Monthly Notices of the Royal Astronomical Society}, \textit{485}(4), 5221-5234.\\

Lodders, K. (2003). Solar system abundances and condensation temperatures of the elements. \textit{The Astrophysical Journal}, \textit{591}(2), 1220.\\

Lubow, S. H., Seibert, M., \& Artymowicz, P. (1999). Disk accretion onto high-mass planets. \textit{The Astrophysical Journal}, \textit{526}(2), 1001.\\

Lunine, J. I., \& Stevenson, D. J. (1982). Formation of the Galilean satellites in a gaseous nebula. \textit{Icarus}, \textit{52}(1), 14-39.\\

Mahaffy, P. R., Donahue, T. M., Atreya, S. K., Owen, T. C., \& Niemann, H. B. (1998). Galileo probe measurements of D/H and 3He/4He in Jupiter's atmosphere. \textit{Space Science Reviews}, \textit{84}(1), 251-263.\\

Martin, R. G., \& Lubow, S. H. (2011). Tidal truncation of circumplanetary discs. \textit{Monthly Notices of the Royal Astronomical Society}, \textit{413}(2), 1447-1461.\\

Meier, R., Owen, T. C., Matthews, H. E., Jewitt, D. C., Bockelee-Morvan, D., Biver, N., ... \& Gautier, D. (1998). A determination of the HDO/H2O ratio in comet C/1995 O1 (Hale-Bopp). \textit{Science}, \textit{279}(5352), 842-844.\\

Mohanty, S., \& Shu, F. H. (2008). Magnetocentrifugally driven flows from young stars and disks. VI. Accretion with a multipole stellar field. \textit{The Astrophysical Journal}, \textit{687}(2), 1323.\\

Morbidelli, A., Szulágyi, J., Crida, A., Lega, E., Bitsch, B., Tanigawa, T., \& Kanagawa, K. (2014). Meridional circulation of gas into gaps opened by giant planets in three-dimensional low-viscosity disks. \textit{Icarus}, \textit{232}, 266-270.\\

Morbidelli, A., Baillie, K., Batygin, K., Charnoz, S., Guillot, T., Rubie, D. C., \& Kleine, T. (2022). Contemporary formation of early Solar System planetesimals at two distinct radial locations. \textit{Nature Astronomy}, \textit{6}(1), 72-79.\\

Moses, J. I. (2014). Chemical kinetics on extrasolar planets. \textit{Philosophical Transactions of the Royal Society A: Mathematical, Physical and Engineering Sciences}, \textit{372}(2014), 20130073.\\

Mosqueira, I., \& Estrada, P. R. (2003). Formation of the regular satellites of giant planets in an extended gaseous nebula I: subnebula model and accretion of satellites. \textit{Icarus}, \textit{163}(1), 198-231.\\

Mousis, O., Gautier, D., \& Bockelée-Morvan, D. (2002). An evolutionary turbulent model of Saturn's subnebula: implications for the origin of the atmosphere of Titan. \textit{Icarus}, \textit{156}(1), 162-175.\\

Mousis, O., \& Gautier, D. (2004). Constraints on the presence of volatiles in Ganymede and Callisto from an evolutionary turbulent model of the Jovian subnebula. \textit{Planetary and Space Science}, \textit{52}(5-6), 361-370.\\

Mousis, O., Schneeberger, A., Lunine, J. I., Glein, C. R., Bouquet, A., \& Vance, S. D. (2023). Early stages of Galilean moon formation in a water-depleted environment. \textit{The Astrophysical journal letters}, \textit{944}(2), L37.\\

Müller, J., Savvidou, S., \& Bitsch, B. (2021). The water-ice line as a birthplace of planets: implications of a species-dependent dust fragmentation threshold. \textit{Astronomy \& Astrophysics}, \textit{650}, A185.\\

Nakagawa, Y., Sekiya, M., \& Hayashi, C. (1986). Settling and growth of dust particles in a laminar phase of a low-mass solar nebula. \textit{Icarus}, \textit{67}(3), 375-390.\\

Nelson, R. P., Gressel, O., \& Umurhan, O. M. (2013). Linear and non-linear evolution of the vertical shear instability in accretion discs. \textit{Monthly Notices of the Royal Astronomical Society}, \textit{435}(3), 2610-2632.\\

Nixon, C. A., Temelso, B., Vinatier, S., Teanby, N. A., Bézard, B., Achterberg, R. K., ... \& Flasar, F. M. (2012). Isotopic ratios in Titan's methane: measurements and modeling. \textit{The Astrophysical Journal}, \textit{749}(2), 159.\\

Ormel, C. W., \& Cuzzi, J. N. (2007). Closed-form expressions for particle relative velocities induced by turbulence. \textit{Astronomy \& Astrophysics}, \textit{466}(2), 413-420.\\

Ostertag, D., \& Flock, M. (2025). Strong clumping in global streaming instability simulations with a dusty fluid. \textit{Astronomy \& Astrophysics}, \textit{695}, L13.\\

Ostriker, E. C., \& Shu, F. H. (1995). Magnetocentrifugally driven flows from young stars and disks. IV. The accretion funnel and dead zone. \textit{Astrophysical Journal v. 447, p. 813}, \textit{447}, 813.\\

Owen, J. E. (2014). Snow lines as probes of turbulent diffusion in protoplanetary disks. \textit{The Astrophysical Journal Letters}, \textit{790}(1), L7.\\

Peale, S. J., \& Lee, M. H. (2002). A primordial origin of the Laplace relation among the Galilean satellites. \textit{Science}, \textit{298}(5593), 593-597.\\

Prinn, R. G., \& Barshay, S. S. (1977). Carbon monoxide on Jupiter and implications for atmospheric convection. \textit{Science}, \textit{198}(4321), 1031-1034.\\

Prinn, R. G., \& Fegley Jr, B. (1981). Kinetic inhibition of CO and N2 reduction in circumplanetary nebulae-Implications for satellite composition. \textit{Astrophysical Journal, Part 1, vol. 249, Oct. 1, 1981, p. 308-317.}, \textit{249}, 308-317.\\

Robert, F., Gautier, D., \& Dubrulle, B. (2000). The solar system D/H ratio: observations and theories. \textit{Space Science Reviews}, \textit{92}(1), 201-224.\\

Ros, K., \& Johansen, A. (2013). Ice condensation as a planet formation mechanism. \textit{Astronomy \& Astrophysics}, \textit{552}, A137.\\

Rosotti, G. P. (2023). Empirical constraints on turbulence in proto-planetary discs. \textit{New Astronomy Reviews}, \textit{96}, 101674.\\

Sasaki, T., Stewart, G. R., \& Ida, S. (2010). Origin of the different architectures of the Jovian and Saturnian satellite systems. \textit{The Astrophysical Journal}, \textit{714}(2), 1052.\\

Schoonenberg, D., \& Ormel, C. W. (2017). Planetesimal formation near the snowline: in or out?. \textit{Astronomy \& Astrophysics}, \textit{602}, A21.\\

Schreiber, A., \& Klahr, H. (2018). Azimuthal and vertical streaming instability at high dust-to-gas ratios and on the scales of planetesimal formation. \textit{The Astrophysical Journal}, \textit{861}(1), 47.\\

Schubert, G., Stevenson, D. J., \& Ellsworth, K. (1981). Internal structures of the Galilean satellites. \textit{Icarus}, \textit{47}(1), 46-59.\\

Sekiya, M. (1998). Quasi-equilibrium density distributions of small dust aggregations in the solar nebula. \textit{Icarus}, \textit{133}(2), 298-309.\\

Shakura, N. I., \& Sunyaev, R. A. (1973). Black holes in binary systems. Observational appearance. \textit{Astronomy and Astrophysics, Vol. 24, p. 337-355}, \textit{24}, 337-355.\\

Shibaike, Y., Ormel, C. W., Ida, S., Okuzumi, S., \& Sasaki, T. (2019). The Galilean satellites formed slowly from pebbles. \textit{The Astrophysical Journal}, \textit{885}(1), 79.\\

Smith, B. A., Soderblom, L. A., Beebe, R., Boyce, J., Briggs, G., Carr, M., ... \& Veverka, J. (1979). The Galilean satellites and Jupiter: Voyager 2 imaging science results. \textit{Science}, \textit{206}(4421), 927-950.\\

Sohl, F., Spohn, T., Breuer, D., \& Nagel, K. (2002). Implications from Galileo observations on the interior structure and chemistry of the Galilean satellites. \textit{Icarus}, \textit{157}(1), 104-119.\\

Sohl, F., Hussmann, H., Schwentker, B., Spohn, T., \& Lorenz, R. D. (2003). Interior structure models and tidal Love numbers of Titan. \textit{Journal of Geophysical Research: Planets}, \textit{108}(E12).\\

Solomon, P. M., \& Woolf, N. J. (1973). Interstellar deuterium: Chemical fractionation. \textit{Astrophysical Journal, vol. 180, p. L89}, \textit{180}, L89.\\

Squire, J., \& Hopkins, P. F. (2018). Resonant drag instabilities in protoplanetary discs: the streaming instability and new, faster growing instabilities. \textit{Monthly Notices of the Royal Astronomical Society}, \textit{477}(4), 5011-5040.\\

Stevenson, D. J., \& Lunine, J. I. (1988). Rapid formation of Jupiter by diffusive redistribution of water vapor in the solar nebula. \textit{Icarus}, \textit{75}(1), 146-155.\\

Suetsugu, R., Ohtsuki, K., \& Fujita, T. (2016). Orbital characteristics of planetesimals captured by circumplanetary gas disks. \textit{The Astronomical Journal}, \textit{151}(6), 140.\\

Suetsugu, R., \& Ohtsuki, K. (2017). Distribution of captured planetesimals in circumplanetary gas disks and implications for accretion of regular satellites. \textit{The Astrophysical Journal}, \textit{839}(1), 66.\\

Szulágyi, J. (2017). Effects of the planetary temperature on the circumplanetary disk and on the gap. \textit{The Astrophysical Journal}, \textit{842}(2), 103.\\

Szulágyi, J., Binkert, F., \& Surville, C. (2022). Meridional circulation of dust and gas in the circumstellar disk: delivery of solids onto the circumplanetary region. \textit{The Astrophysical Journal}, \textit{924}(1), 1.\\

Takahashi, S. Z., \& Inutsuka, S. I. (2014). Two-component secular gravitational instability in a protoplanetary disk: a possible mechanism for creating ring-like structures. \textit{The Astrophysical Journal}, \textit{794}(1), 55.\\

Taki, T., Fujimoto, M., \& Ida, S. (2016). Dust and gas density evolution at a radial pressure bump in protoplanetary disks. \textit{Astronomy \& Astrophysics}, \textit{591}, A86.\\

Tanigawa, T., Ohtsuki, K., \& Machida, M. N. (2012). Distribution of accreting gas and angular momentum onto circumplanetary disks. \textit{The Astrophysical Journal}, \textit{747}(1), 47.\\

Tanigawa, T., Maruta, A., \& Machida, M. N. (2014). Accretion of solid materials onto circumplanetary disks from protoplanetary disks. \textit{The Astrophysical Journal}, \textit{784}(2), 109.\\

Teague, R., Bae, J., \& Bergin, E. A. (2019). Meridional flows in the disk around a young star. \textit{Nature}, \textit{574}(7778), 378-381.\\

Toomre, A. (1964). On the gravitational stability of a disk of stars. \textit{Astrophysical Journal, vol. 139, p. 1217-1238 (1964).}, \textit{139}, 1217-1238.\\

Tyler, G. L., Eshleman, V. R., Anderson, J. D., Levy, G. S., Lindal, G. F., Wood, G. E., \& Croft, T. A. (1981). Radio science investigations of the Saturn system with Voyager 1: Preliminary results. \textit{Science}, \textit{212}(4491), 201-206.\\

Waite Jr, J. H., Lewis, W. S., Magee, B. A., Lunine, J. I., McKinnon, W. B., Glein, C. R., ... \& Ip, W. H. (2009). Liquid water on Enceladus from observations of ammonia and 40Ar in the plume. \textit{Nature}, \textit{460}(7254), 487-490.\\

Wang, D., Lunine, J. I., \& Mousis, O. (2016). Modeling the disequilibrium species for Jupiter and Saturn: Implications for Juno and Saturn entry probe. \textit{Icarus}, \textit{276}, 21-38.\\

Weiss, L. M., Marcy, G. W., Petigura, E. A., Fulton, B. J., Howard, A. W., Winn, J. N., ... \& Cargile, P. A. (2018). The California-Kepler survey. V. Peas in a pod: Planets in a Kepler multi-planet system are similar in size and regularly spaced. \textit{The Astronomical Journal}, \textit{155}(1), 48.\\

Yang, L., Ciesla, F. J., \& Alexander, C. M. D. (2013). The D/H ratio of water in the solar nebula during its formation and evolution. \textit{Icarus}, \textit{226}(1), 256-267.\\

Yap, T. E., \& Batygin, K. (2024). Dust-gas coupling in turbulence-and MHD wind-driven protoplanetary disks: Implications for rocky planet formation. \textit{Icarus}, \textit{417}, 116085.\\

Yap, T. E., Batygin, K., \& Tissot, F. L. (2025). Early Solar System Turbulence Constrained by High Oxidation States in the Oldest Noncarbonaceous Planetesimals. \textit{The Planetary Science Journal}, \textit{6}(1), 2.\\

Yap, T. E., \& Batygin, K. (2025). Callisto’s Non-Resonant Orbits as an Outcome of Circum-Jovian Disk Substructure. \textit{The Astrophysical Journal. (In Rev.)} \\

Youdin, A. N., \& Shu, F. H. (2002). Planetesimal formation by gravitational instability. \textit{The Astrophysical Journal}, \textit{580}(1), 494.\\

Youdin, A. N., \& Goodman, J. (2005). Streaming instabilities in protoplanetary disks. \textit{The Astrophysical Journal}, \textit{620}(1), 459.\\

Youdin, A. N., \& Lithwick, Y. (2007). Particle stirring in turbulent gas disks: Including orbital oscillations. \textit{icarus}, \textit{192}(2), 588-604.\\

\end{document}